\documentclass{article}

\usepackage{arxiv}

\usepackage[utf8]{inputenc}
\usepackage[T1]{fontenc}
\usepackage[hypertexnames=false]{hyperref}
\usepackage{url}
\usepackage{booktabs}
\usepackage{amsfonts}
\usepackage{amsmath,amssymb,amsfonts}
\usepackage{amsthm}
\usepackage{mathrsfs}
\usepackage[title]{appendix}
\usepackage{xcolor}
\usepackage{textcomp}
\usepackage{manyfoot}
\usepackage{algorithm}
\usepackage{algorithmicx}
\usepackage{algpseudocode}
\usepackage{listings}
\usepackage{siunitx}
\usepackage{longtable}
\usepackage{array}
\usepackage{microtype}
\usepackage{graphicx}
\usepackage{placeins}
\graphicspath{{./img/}{./images/}}

\title{Qronecker: A Certifiable Kronecker Compression Primitive for Quantum-Chemistry Hamiltonians}

\author{
  Yuqi Zhang \\
  Department of Computer Science \\
  Kent State University \\
  Kent, OH 44240, USA \\
  Cleveland Clinic Genome Center \\
  Cleveland Clinic \\
  Cleveland, OH 44106, USA \\
  \And
  Sixu Chen \\
  Department of Computer Science \\
  Kent State University \\
  Kent, OH 44240, USA \\
  \And
  Feixiong Cheng \\
  Cleveland Clinic Genome Center \\
  Cleveland Clinic \\
  Cleveland, OH 44106, USA \\
  \And
  Qiang Guan \\
  Department of Computer Science \\
  Kent State University \\
  Kent, OH 44240, USA \\
  \texttt{qguan@kent.edu} \\
}

\date{}

\begin{document}
\maketitle

\begin{abstract}
Processing qubit Hamiltonians derived from electronic-structure problems can become classically prohibitive because many downstream manipulations still rely on dense operator constructions whose cost grows exponentially with qubit number. We introduce Qronecker, a cut-aware low-rank Kronecker decomposition algorithm that turns Hamiltonian compression into a certifiable, resource-aware decision primitive. Operating entirely in Pauli coefficient space, Qronecker avoids forming dense $2^n \times 2^n$ matrices, constructs low-rank Kronecker approximations under a chosen bipartition, and returns both an instance-specific compressibility curve and a state-independent worst-case energy certificate that links rank and cut choices to conservative energy-deviation bounds. Across molecular benchmarks comprising hundreds of systems up to 30 qubits, we find that traceless low-rank structure is common but heterogeneous: many screened systems reach high coefficient-space fidelity at low rank, yielding large savings in classical preprocessing and conditional reductions in downstream circuit-resource proxies, while the certificate remains valid but conservative on the auditable subset. The same analysis shows that fixed global fidelity targets are not generally sufficient for chemistry-level guarantees, motivating adaptive rank and cut selection. These results position Qronecker as a certifiable compression primitive for rank and cut selection in quantum-chemistry Hamiltonian processing.
\end{abstract}

\keywords{Quantum chemistry \and Hamiltonian simulation \and Tensor decomposition \and Quantum computing \and Computational scalability}

\section{Introduction}

Electronic-structure Hamiltonians are central objects in quantum chemistry and constitute the fundamental input to quantum algorithms for predicting molecular energies and properties. After mapping a fermionic model to qubits, the Hamiltonian is naturally expressed as a Pauli-sum operator acting on $n$ qubits; however, many practical workflows still rely on dense operator constructions or dense linear-algebra routines, so classical memory and compute costs can grow rapidly with system size before quantum execution becomes the limiting factor~\cite{danilov2025enhancing,robledo2025chemistry}. This challenge is particularly pronounced in hybrid workflows such as Sample-based Quantum Diagonalization (SQD)~\cite{robledo2025chemistry,sugisaki2025hamiltonian}, where quantum hardware generates samples but Hamiltonian projection, matrix-element evaluation, and iterative diagonalization remain classical tasks. Similar bottlenecks arise earlier in quantum-chemistry pipelines during operator assembly, mapping, and bookkeeping. End-to-end scalability therefore depends not only on the chemistry model, but also on whether Hamiltonian structure can be converted into explicit, auditable resource decisions without dense materialization.

Existing approaches for compression and cost reduction broadly fall into two categories. One class exploits physical structure by reducing effective degrees of freedom through approximations such as active-space selection~\cite{stein2016automated} and frozen-core treatments~\cite{yu2021accurate}. A second class seeks to preserve the underlying physical model while reducing the cost of Hamiltonian manipulation through representational and computational structure. Representative strategies include exploiting low-rank or sparse structure at the integral and two-body level~\cite{motta2021low,weigend2009approximated}, leveraging orbital locality~\cite{mcclean2014exploiting}, using structured operator or tensor representations~\cite{hohenstein2012tensor,kolda2009tensor}, and lowering measurement cost through Pauli-term grouping~\cite{verteletskyi2020measurement}. These methods mitigate important bottlenecks, but two requirements remain insufficiently coupled to deployment. First, compressibility is strongly instance dependent, so practical use requires reliable screening to determine when compression is effective for a given molecule, modeling setting, and subsystem partition. Second, the compression level must be tied to computable conservative error guarantees so that rank and cut selection can be made under explicit accuracy constraints rather than by heuristic thresholding alone.

To address this gap, we introduce Qronecker---a cut-aware low-rank Kronecker decomposition algorithm that also serves as a certifiable, resource-aware decision primitive for scalable processing of quantum-chemistry Hamiltonians. Without constructing dense $2^n \times 2^n$ matrices, Qronecker reshapes the traceless Pauli coefficients of a Hamiltonian into a cut-dependent matrix and extracts a low-rank Kronecker representation under a chosen bipartition $A\mid B$~\cite{van1993approximation,eckart1936approximation,kolda2009tensor}. The method returns an instance-specific compressibility curve $\rho_k$ together with a conservative energy certificate,
\begin{equation}
\Delta E_{\mathrm{bound}}(k)\le \|H_{\mathrm{tr}}\|_F\sqrt{1-\rho_k},
\end{equation}
which converts cut and rank from heuristic hyperparameters into auditable accuracy--resource decision variables. In this formulation, compression is evaluated against the same residual quantity that later controls the certificate, and the same coefficient-space statistics support screening, rank selection, resource accounting, and retention of the reference treatment when necessary. The bound also makes clear that the adequacy of any fixed global $\rho$ target is system dependent through $\|H_{\mathrm{tr}}\|_F$, so chemical-accuracy deployment requires adaptive rank policies rather than a universal truncation rule.

The empirical study follows the same progression that motivates the method. We first quantify the prevalence and geometry robustness of low-rank structure on a boundary-scan cohort of $765$ molecular Hamiltonians. We then analyze rank profiles, cut sensitivity, classical preprocessing gains, and circuit-resource proxies on a performance-test cohort of $400$ systems. Finally, we audit the certificate on the size-feasible subset with reference energies and run a separate chemistry-accuracy boundary test on a fixed set of twenty $12$-qubit cases. The results show that traceless low-rank structure is widespread but heterogeneous, that most screened systems reach high coefficient-space fidelity at low rank with large classical-side gains, and that the certificate is valid but substantially conservative. They also show that high-fidelity truncation and worst-case chemical certification are distinct operating regimes: ranks sufficient for large resource savings are often far below the ranks required to certify chemical accuracy.

In hybrid quantum-chemistry computation, Qronecker can serve as an operator-preprocessing layer that reduces computational burden across multiple stages. In SQD-style workflows, for example, a cut-aware low-rank Kronecker representation can reduce the cost of matrix-element evaluation, projection construction, and related linear-algebra operations, while $\rho_k$ and the certificate provide explicit constraints for rank selection and retention of the reference treatment when required. In this work, we focus on validating the primitive itself and its decision interface rather than on optimizing a single end-to-end application. The resulting picture is a systematic procedure in which systems are screened, cuts and ranks are selected against explicit structural or accuracy targets, and instances outside the certified regime are left to the reference method.

The main contributions of this work are:
(1) We propose Qronecker, a cut-aware low-rank Kronecker decomposition algorithm for Pauli-sum quantum-chemistry Hamiltonians that operates in coefficient space without constructing dense operators, thereby reducing the memory and computational cost of Hamiltonian processing.
(2) We formulate the state-independent worst-case energy certificate specialized to coefficient-space Kronecker compression of Pauli-sum quantum-chemistry Hamiltonians, which turns the compressibility curve $\rho_k$ into a direct criterion for screening, rank selection, and chemistry-feasibility auditing.
(3) We provide a certifiable decision interface in which the instance-specific $\rho_k$ curve and the conservative energy certificate convert cut and rank into auditable accuracy--resource variables for screening, target selection, and retention of the reference treatment when necessary.
(4) Through boundary screening, rank/resource benchmarking, certificate audit, and chemistry-boundary testing, we show that low-rank coefficient-space structure is common but heterogeneous, that it yields substantial classical-side savings and conditional downstream circuit benefits, and that the method supports a continuum of operating points from high-fidelity approximation to chemically certified compression on the auditable cases studied. In this regime, chemistry-level guarantees remain attainable, but generally require higher ranks and an explicit accuracy--resource trade-off rather than fixed global fidelity targets such as $\rho=0.999$ or $\rho=0.9999$.

\section{Results}
\label{sec:results}

\subsection{Coefficient-Space Kronecker Formulation, Metrics, and Certificates}
\label{subsec:results_formulation}

We first fix the notation used throughout this section. All screening, rank selection, resource accounting, and certificate evaluation are carried out in coefficient space after removing the identity offset, so the same quantities support every later claim.

Let an $n$-qubit Hamiltonian be represented as a Pauli-sum operator
\begin{equation}
H=\sum_{p\in\mathcal{P}_n} c_p\,P_p,\qquad P_p\in\{I,X,Y,Z\}^{\otimes n},\ \ c_p\in\mathbb{R},
\end{equation}
where $\mathcal{P}_n$ indexes $n$-qubit Pauli strings. We separate the identity contribution and perform all compressibility and certificate accounting on the traceless part,
\begin{equation}
c_0 \equiv c_{I^{\otimes n}},\qquad H_{\mathrm{tr}} \equiv H - c_0 I.
\end{equation}

\paragraph{Cut-aware coefficient matrix and rank-$k$ Kronecker form.}
Fix a bipartition $A\mid B$ with $n_A$ qubits in $A$ and $n_B=n-n_A$ qubits in $B$. Any Pauli string factorizes as $P_p=P_a\otimes P_b$ with $P_a\in\{I,X,Y,Z\}^{\otimes n_A}$ and $P_b\in\{I,X,Y,Z\}^{\otimes n_B}$. We reshape the traceless coefficients into the cut-dependent matrix
\begin{equation}
C\in\mathbb{R}^{4^{n_A}\times 4^{n_B}},\qquad C_{a,b}\equiv c_{(a,b)}\ \ \text{for}\ \ P_{(a,b)}=P_a\otimes P_b,
\end{equation}
which is the object decomposed by Qronecker without ever constructing the dense $2^n\times 2^n$ operator. Let $\sigma_1\ge\sigma_2\ge\cdots$ be the singular values of $C$. The rank-$k$ truncation
\begin{equation}
C_k \equiv \sum_{r=1}^{k}\sigma_r u_r v_r^{\top}
\end{equation}
induces a rank-$k$ Kronecker approximation of the traceless Hamiltonian,
\begin{equation}
\widetilde{H}_{k,\mathrm{tr}} \equiv \sum_{r=1}^{k}\alpha_r\big(A_r\otimes B_r\big),\qquad
\widetilde{H}_k \equiv c_0 I + \widetilde{H}_{k,\mathrm{tr}},
\end{equation}
where $A_r$ and $B_r$ are subsystem Pauli-sum operators encoded by $u_r$ and $v_r$ under a fixed basis ordering.

\paragraph{Compressibility metrics.}
Instance-wise compressibility is quantified by the cumulative captured energy in coefficient space,
\begin{equation}
\rho_k \equiv \frac{\|C_k\|_F^2}{\|C\|_F^2}
= \frac{\sum_{i=1}^{k}\sigma_i^2}{\sum_{i\ge 1}\sigma_i^2}\in[0,1],
\qquad
\rho_{1,\mathrm{tr}}\equiv \rho_1\ \ \text{computed from}\ \ H_{\mathrm{tr}}.
\end{equation}
We also use the marginal gain
\begin{equation}
\Delta\rho_k \equiv \rho_k-\rho_{k-1},\qquad \rho_0\equiv 0,
\end{equation}
and, for a target fidelity threshold $\tau\in(0,1)$, the first-hit rank
\begin{equation}
k^{\star}(\tau)\equiv \min\{k:\rho_k\ge \tau\},
\end{equation}
treating instances that do not reach $\tau$ by the maximum scanned rank $k_{\max}$ as right-censored at $k_{\max}$.
Unless noted otherwise, every $\rho_k$ reported below is computed on the traceless coefficient matrix $C(H_{\mathrm{tr}};A\mid B)$.

\paragraph{Certificate derivation and audit quantities.}
Because the Pauli basis satisfies $\mathrm{Tr}(P_pP_q)=2^n\delta_{pq}$, the Frobenius norm of the traceless Hamiltonian is
\begin{equation}
\|H_{\mathrm{tr}}\|_F^2 = 2^n\|C\|_F^2.
\end{equation}
The rank-$k$ residual therefore obeys
\begin{equation}
\|C-C_k\|_F=\|C\|_F\sqrt{1-\rho_k}
\quad\Longrightarrow\quad
\|H_{\mathrm{tr}}-\widetilde{H}_{k,\mathrm{tr}}\|_F=\|H_{\mathrm{tr}}\|_F\sqrt{1-\rho_k}~\cite{johnson1963theorem}.
\end{equation}
Using $\|X\|_2\le \|X\|_F$ together with the standard eigenvalue perturbation bound $|E_0(H)-E_0(\widetilde{H}_k)|\le \|H-\widetilde{H}_k\|_2$, we obtain the conservative worst-case energy certificate
\begin{equation}
\Delta E_{\mathrm{bound}}(k)\ \equiv\ \|H_{\mathrm{tr}}-\widetilde{H}_{k,\mathrm{tr}}\|_2
\ \le\ \|H_{\mathrm{tr}}\|_F\sqrt{1-\rho_k}~\cite{weyl1912asymptotische}.
\label{eq:energy_certificate_results}
\end{equation}
When reference energies are available, we audit validity and conservatism through the tightness ratio
\begin{equation}
\eta(k)\equiv \frac{|E_0(H)-E_0(\widetilde{H}_k)|}{\Delta E_{\mathrm{bound}}(k)}.
\end{equation}
For chemistry-oriented discussion we use $\epsilon_{\mathrm{chem}}=1$ kcal/mol $\approx 1.5936\times 10^{-3}$ Ha and the induced instance-dependent requirement
\begin{equation}
\rho_{\mathrm{required}} \equiv 1-\left(\frac{\epsilon_{\mathrm{chem}}}{\|H_{\mathrm{tr}}\|_F}\right)^2,
\end{equation}
which follows by enforcing $\Delta E_{\mathrm{bound}}(k)\le \epsilon_{\mathrm{chem}}$ in Eq.~\eqref{eq:energy_certificate_results}.

\paragraph{Resource metrics.}
The same formulation also implies the preprocessing-cost scaling of
coefficient-space compression. For
$C\in\mathbb{R}^{4^{n_A}\times 4^{n_B}}$, a dense
SVD~\cite{golub1971singular} requires time on the order of
$\mathcal{O}(4^n\cdot 4^{\min(n_A,n_B)})$ and memory $\Theta(4^n)$ to
materialize $C$, whereas Qronecker targets only the leading $k$
components and can operate through sparse or implicit matrix--vector
products. We report realized benefits with ratio metrics (values larger
than $1$ indicate improvement):
\begin{equation}
\mathrm{Speedup}\equiv \frac{t_{\mathrm{dense}}}{t_{\mathrm{decomp}}},\qquad
\mathrm{Memory\ Ratio}\equiv \frac{\mathrm{bytes}_{\mathrm{dense}}}{\mathrm{bytes}_{\mathrm{decomp}}},
\end{equation}
\begin{equation}
\mathrm{Depth\ Gain}\equiv \frac{\mathrm{depth}_{\mathrm{full}}}{\mathrm{depth}_{\mathrm{cut,par}}},\qquad
\mathrm{2Q\ Gain}\equiv \frac{\#2\mathrm{Q}_{\mathrm{full}}}{\#2\mathrm{Q}_{\mathrm{cut}}}.
\end{equation}
With these definitions fixed, the rest of the section follows the study design introduced above: we first screen for low-rank structure, then translate that structure into rank budgets, next measure the realized resource gains, and finally audit the conservative energy guarantee.

\subsection{Boundary Screening and Traceless Compressibility}
\label{subsec:results_boundary}

We begin by fixing the empirical scope of the screening problem. Results are reported on two chemistry cohorts: a boundary-scan dataset (BS, $N=765$), which densely samples an interatomic distance parameter $R$ when that parameter is recoverable from the system name, and a performance-test dataset (PT, $N=400$), which is used for rank, resource, and certificate evaluation. Both cohorts extend to $n=30$ qubits (BS: $n\in[2,30]$ with median $22$; PT: $n\in[12,30]$ with median $24$) and are dominated by STO-3G instances~\cite{hehre1969self}; Appendix Tables~\ref{tab:dataset_boundary_overview} and~\ref{tab:dataset_performance_overview} summarize the exact cohort composition. When $R$ is detectable, BS contains $101$ scan series (median $4$ points per series, maximum $10$) over $R\in[0.50,10.80]$, while PT contains $77$ shorter series (median $4$, maximum $5$) over $R\in[0.75,10.80]$.

\begin{figure*}[htbp]
    \centering
    \includegraphics[width=\textwidth]{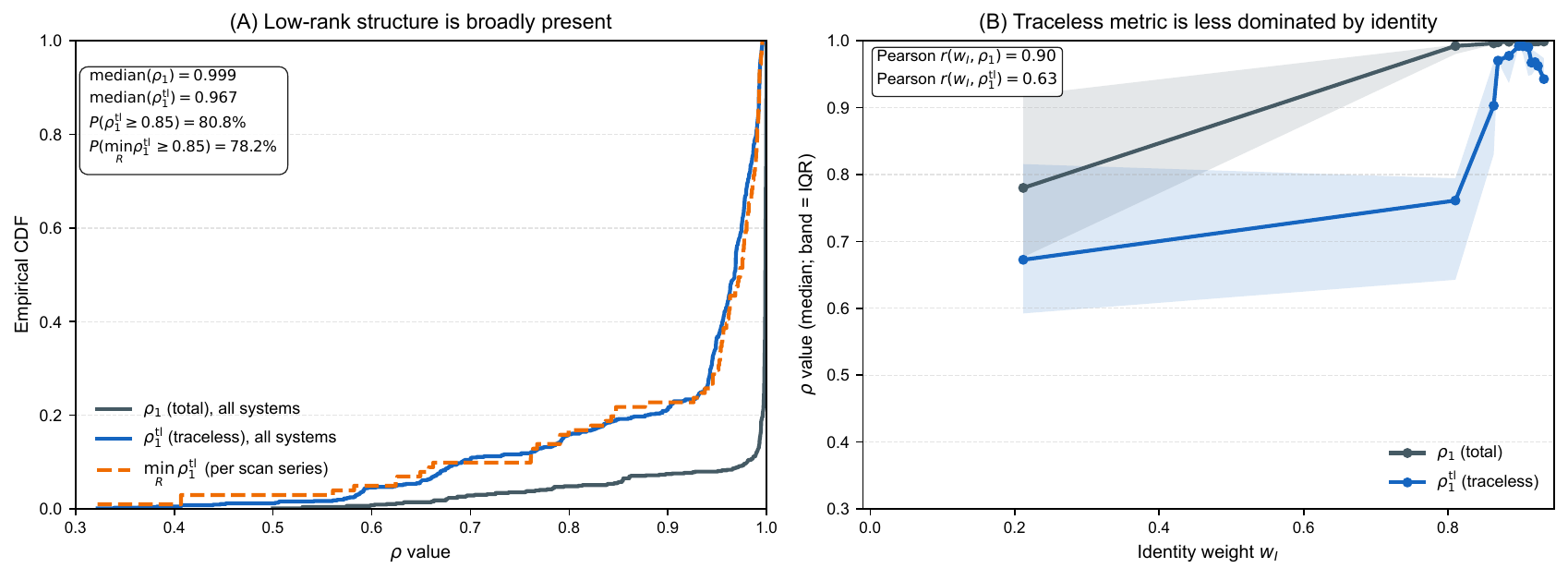}
    \caption{Boundary screening on the boundary-scan cohort. (A) Empirical CDFs of rank-1 capture for the total-space metric $\rho^{\mathrm{tot}}_1$, the traceless metric $\rho_{1,\mathrm{tr}}$, and the per-scan-series worst case $\min_R\rho_{1,\mathrm{tr}}$. (B) Dependence of $\rho^{\mathrm{tot}}_1$ and $\rho_{1,\mathrm{tr}}$ on the identity weight $w_I$, illustrating that the total-space metric is strongly confounded by identity contributions.}
    \label{fig:lowrank_universality}
\end{figure*}

Figure~\ref{fig:lowrank_universality}(A) shows that rank-1 traceless compressibility is prevalent on BS, but varies substantially across instances. The median screening signal is $\mathrm{median}(\rho_{1,\mathrm{tr}})=0.967$, and $80.8\%$ of instances satisfy $\rho_{1,\mathrm{tr}}\ge 0.85$ under the reported cut settings. The fraction below $0.85$ increases markedly on the stress cohort: the common cohort inputs ($N=685$) have $\mathrm{median}(\rho_{1,\mathrm{tr}})=0.969$ and only $15.3\%$ of systems fall below $0.85$, whereas the stress cohort inputs ($N=80$) shift to $\mathrm{median}(\rho_{1,\mathrm{tr}})=0.839$ with $52.5\%$ below $0.85$. These statistics indicate that low-rank structure is widespread but not uniform, so explicit screening remains necessary.

This distinction also affects later rank requirements. Systems with smaller $\rho_{1,\mathrm{tr}}$ are less compatible with a rank-1 approximation and, in the later rank scans, typically require larger rank budgets to reach a fixed fidelity target. Geometry scans exhibit the same behavior. Among the $101$ detectable-$R$ series in BS, $78.2\%$ satisfy the conservative series-level criterion $\min_R \rho_{1,\mathrm{tr}} \ge 0.85$. Under this rule, a single low-compressibility geometry determines the screening status of the entire scan series. Screening is therefore applied instance-wise and, for geometry scans, geometry-wise.

Figure~\ref{fig:lowrank_universality}(B) quantifies the effect of identity removal on the screening metric. On BS, the total-space indicator $\rho^{\mathrm{tot}}_1$ is almost always close to one ($\mathrm{median}(\rho^{\mathrm{tot}}_1)=0.999$) and is strongly correlated with the identity-component weight ($\mathrm{Pearson}\ r(w_I,\rho^{\mathrm{tot}}_1)=0.90$). This indicates that the total-space metric can be dominated by the identity offset. Removing that offset materially reduces the confounding ($\mathrm{Pearson}\ r(w_I,\rho_{1,\mathrm{tr}})=0.63$) and aligns the screening signal with the residual term that appears in Eq.~\eqref{eq:energy_certificate_results}. The traceless metric is therefore used throughout the remainder of the section.

Compressibility is also cut-dependent because the reshaped matrix $C(H_{\mathrm{tr}};A\mid B)$ depends on the bipartition. The cut is therefore treated as an explicit experimental factor. Appendix Figure~\ref{fig:cut_sensitivity} shows that both $\rho_1$ and the realized resource ratios vary materially across balanced and unbalanced cuts.

\subsection{Rank Profiles and Information Concentration}
\label{subsec:results_rank_profiles}

\begin{figure*}[htbp]
    \centering
    \includegraphics[width=\textwidth]{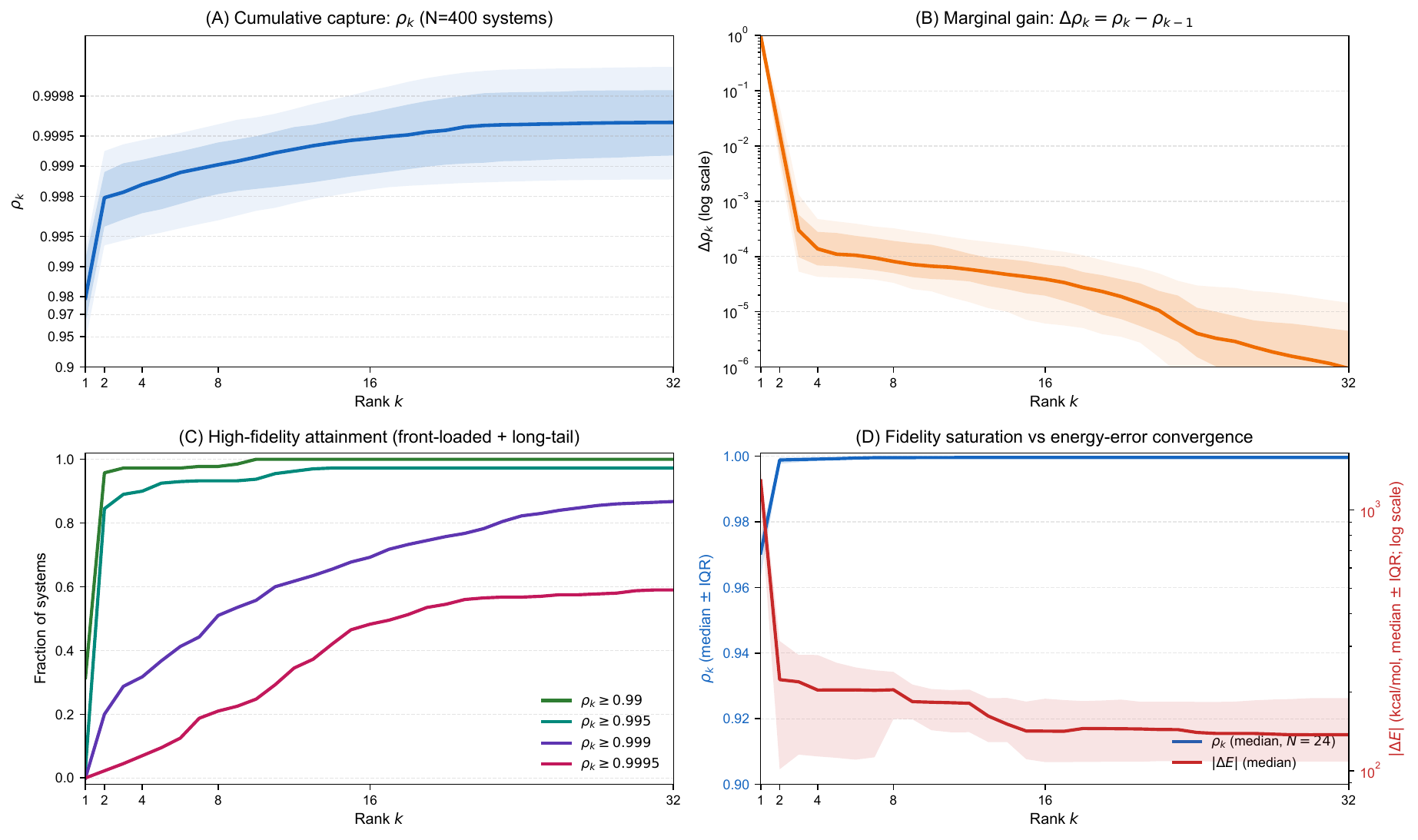}
    \caption{Rank profiles on the performance-test cohort. (A) Cumulative captured energy $\rho_k$ with uncertainty bands. (B) Marginal gains $\Delta\rho_k$ on a log scale, showing rapid decay. (C) Fraction of systems reaching high-fidelity thresholds versus $k$, showing that most successful cases reach the target at low rank while a subset requires higher rank. (D) On the evaluated subset with ground-state errors, $\rho_k$ reaches a high-capture regime earlier than the observed energy error, motivating rank selection under explicit accuracy or certificate constraints.}
    \label{fig:rank_profiles}
\end{figure*}

For screened instances, the rank analysis quantifies how much coefficient-space information is recovered at low rank and how the incremental gain decays as rank increases. On the PT cohort ($N=400$), scanned up to $k_{\max}=32$, Figure~\ref{fig:rank_profiles} shows that capture is concentrated in the first few ranks. The median captured energy rises from $\rho_1=0.979$ at $k=1$ to $0.998$ at $k=2$, reaches $0.9990$ by $k=8$, and only moves to $0.9996$ by $k=32$. The same pattern appears in the marginal gains: the median increment drops from $\Delta\rho_2=1.64\times 10^{-2}$ to $\Delta\rho_4=1.38\times 10^{-4}$ and reaches the $10^{-6}$ scale by $k=32$. The first few terms therefore account for most of the coefficient-space structure, while higher ranks contribute progressively smaller corrections.

This concentration of capture translates directly into rank budgeting. For the target $\tau=0.999$, $347/400$ PT systems reach $\rho_k\ge \tau$ within $k\le 32$, with median required rank $k^{\star}(0.999)=7$ among the reaching systems and a 75th percentile of $14$; the remaining $53$ systems are right-censored at $k_{\max}=32$. The screening signal remains predictive at this stage: after splitting PT into quartiles by $\rho_1$, the lowest quartile has $\mathrm{median}\,k^{\star}(0.999)=15.5$, compared with $7.5$ in the next quartile and $4.0$ in the third quartile. The relation is not deterministic, but it remains monotone, as reflected by the negative Spearman correlation $\rho(\rho_1,k^{\star}(0.999))=-0.32$ on the reaching subset. At the stricter target $\tau=0.9995$, higher-rank requirements become more pronounced: only $236/400$ systems reach the threshold within $k\le 32$, and the median required rank rises to $12$ among the reaching systems.

Figure~\ref{fig:rank_profiles}(D) also indicates that $\rho_k$ alone is insufficient as a rank-selection rule when the downstream objective is energy accuracy. On the $N=24$ systems with observed ground-state errors, $\rho_k$ reaches a high-capture regime at small $k$, but the actual energy error continues to decrease more gradually as rank grows; the median absolute error drops by nearly one order of magnitude from $k=1$ to $k=32$ even though the median $\rho_k$ already exceeds $0.997$ at $k=2$. Appendix Figure~\ref{fig:rank_profiles_addon} shows that this pattern is not limited to median summaries: the per-system heatmaps show systematic concentration of $\Delta\rho_k$ at low rank together with heterogeneous energy-convergence patterns across instances. These results indicate that $\rho_1$ is informative for screening and $\rho_k$ is informative for structural progress, but the final rank choice should still be tied to an application target or to the auditable envelope in Eq.~\eqref{eq:energy_certificate_results}.

\subsection{Resource Gains from Coefficient-Space Decomposition}
\label{subsec:results_resource_gains}

\begin{figure*}[htbp]
    \centering
    \includegraphics[width=\textwidth]{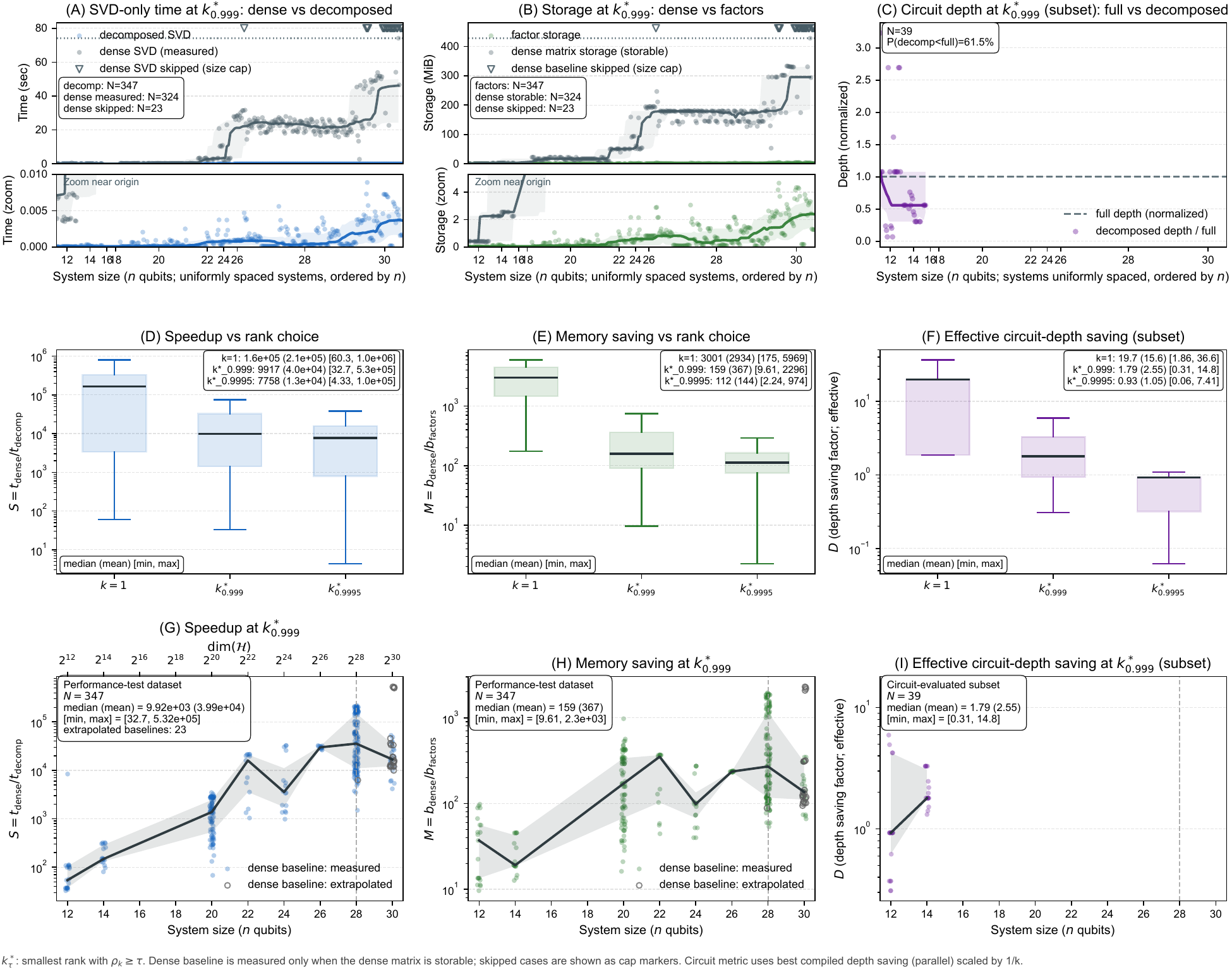}
    \caption{Resource benefits on the performance-test cohort at high-fidelity operating points and their size dependence. Subpanels report dense versus decomposed SVD time and storage at $k^{\star}_{0.999}$, circuit-depth ratios on the circuit-evaluation subset, and distribution comparisons across rank choices ($k=1$ and threshold-selected $k^{\star}$).}
    \label{fig:resource_benefits}
\end{figure*}

\begin{figure*}[htbp]
    \centering
    \includegraphics[width=\textwidth]{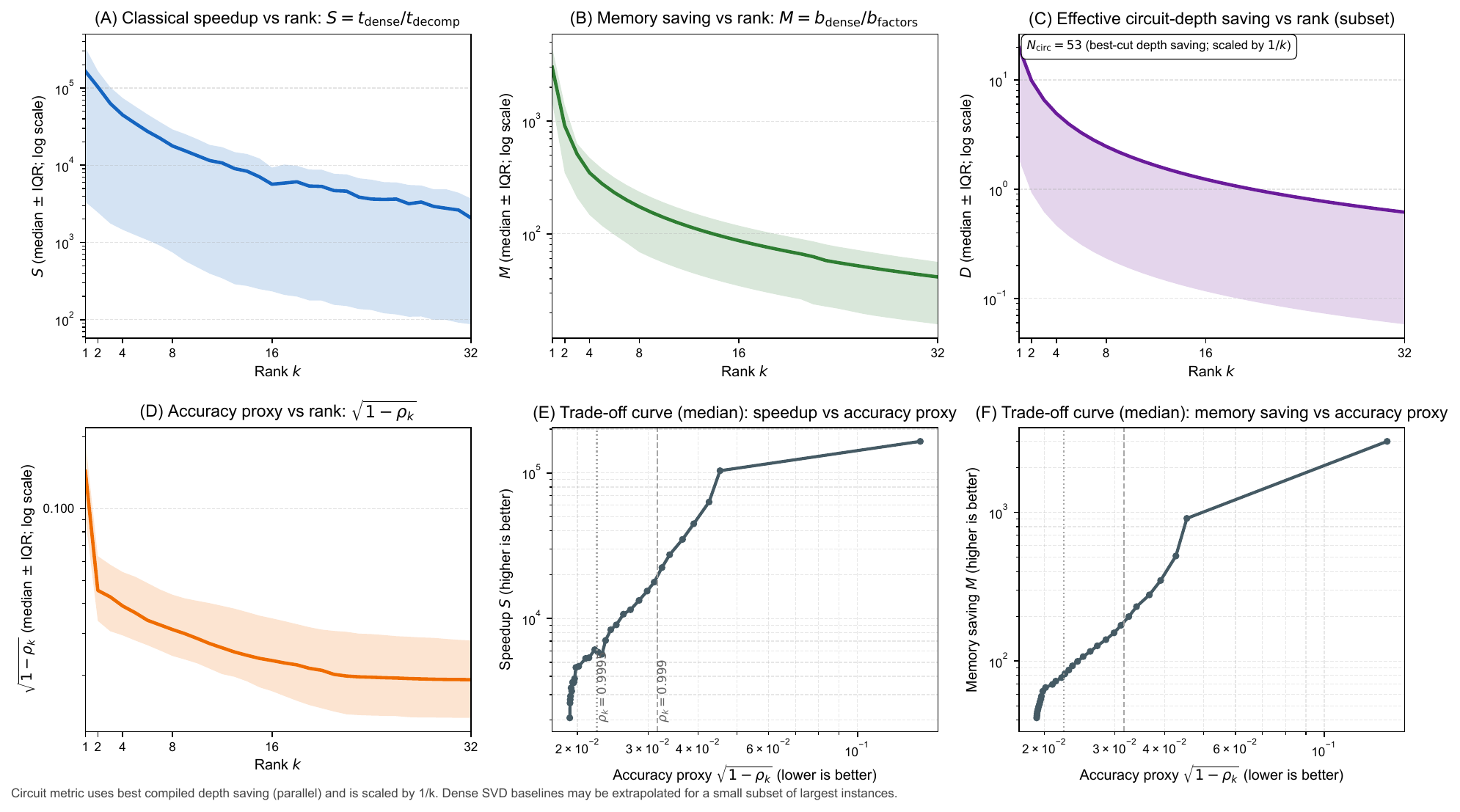}
    \caption{Rank--accuracy--resource trade-off on the performance-test cohort. Panels show how classical speedup and memory saving decrease with rank, how effective circuit-depth saving changes on the circuit-evaluation subset, and how the residual indicator $\sqrt{1-\rho_k}$ improves. Trade-off curves summarize the median relationship between resource factors and this residual indicator across ranks.}
    \label{fig:rank_resource_tradeoff}
\end{figure*}

After fixing the screening and rank-budget rules, we measure the realized gains from using coefficient-space Kronecker decomposition as a preprocessing primitive. Classical benefits are measured against the dense coefficient-matrix SVD baseline, while circuit-side benefits are reported on the subset with matched circuit-evaluation data. Appendix Table~\ref{tab:resource_benefit_overview} summarizes the main operating points; Figures~\ref{fig:resource_benefits} and~\ref{fig:rank_resource_tradeoff} show how those gains vary with size and rank.

At rank $k=1$, the classical benefit is already very large across the full PT cohort. The SVD-only speedup has median $1.65\times 10^{5}$, ranging from $60.3$ to $1.04\times 10^{6}$, and the memory ratio has median $3\times 10^{3}$, ranging from $175$ to $5.97\times 10^{3}$. Moving to stricter fidelity targets reduces these factors but does not remove them. At the first rank satisfying $\rho_k\ge 0.999$, the median speedup remains $9.92\times 10^{3}$ over the $347$ reaching systems and the median memory ratio remains $159$; even at $\rho_k\ge 0.9995$, the median speedup is still $7.76\times 10^{3}$ over $236$ systems and the median memory ratio is $112$. High-fidelity truncations therefore still preserve substantial classical savings for a large portion of the dataset.

Figure~\ref{fig:resource_benefits} shows that these gains increase with system size, corresponding to the regime in which dense preprocessing becomes least practical. At $k=1$, the median speedup reaches $3.05\times 10^{5}$ for $n=28$ and $6.65\times 10^{5}$ for $n=30$. The improvement persists even at the high-fidelity operating point $\rho_k\ge 0.999$: the median speedup is $3.57\times 10^{4}$ at $n=28$ and $1.68\times 10^{4}$ at $n=30$, with corresponding median memory ratios of $270$ and $136$. The dense baseline is increasingly difficult to materialize in this regime and is therefore explicitly marked as measured or extrapolated in the figure, whereas the decomposed path remains executable across the full PT cohort. Over the measured size range, the gap between dense and decomposed preprocessing increases with system size.

This behavior is already visible in the directly benchmarked cases, before any
extrapolation is invoked. Among the PT systems included in the direct dense
comparison, $20$ cases exceeded the prespecified dense-storage threshold; these cases
are all STO-3G and are concentrated in the largest-\(n\) regime (nineteen at $n=30$
and one at $n=28$). The coefficient-space decomposition remained feasible for all of
them. On this subset, the median decomposed SVD time is $5.33$ ms (90th percentile
$8.71$ ms), compared with a median of $0.648$ ms on the PT systems for which direct
dense comparison remained feasible. Once dense materialization reaches the storage
ceiling, the cost of the decomposed route therefore continues to increase only slowly
while remaining computationally tractable.

Rank then controls how much of that benefit is retained. Figure~\ref{fig:rank_resource_tradeoff} makes this trade-off explicit: as $k$ increases, the residual indicator $\sqrt{1-\rho_k}$ decreases monotonically, but both the time speedup and the memory ratio contract. This coupling is also visible on the quantum-facing side. On the circuit-evaluated subset ($N=53$), the effective depth gain has median $19.7$ at $k=1$, falls to $1.79$ on the subset reaching $\rho_k\ge 0.999$ ($N=39$), and further drops to $0.929$ on the subset reaching $\rho_k\ge 0.9995$ ($N=19$). These results show that classical and circuit-side gains decrease at different rates as rank increases, so both should be evaluated at the selected accuracy target.

The cut choice introduces a second decision variable. Appendix Figure~\ref{fig:cut_sensitivity} shows that even an extreme imbalance such as $n_A=2$ still preserves substantial rank-1 capture, with $\mathrm{median}(\rho_1)=0.884$. The realized resource factors nevertheless change materially because both the dense baseline and the factor dimensions are cut-dependent. For example, the median decomposition time increases from $4.9$ ms at $n_A=5$ to $8.7$ ms at $n_A=2$, while the median memory ratio drops from $8.16$ to $0.79$, so an overly imbalanced cut can remove the storage benefit at rank $1$. Cut selection therefore affects speed, memory, and compiled-circuit proxies and should be optimized jointly with the rank budget.

\subsection{Certificate Audit and Chemistry-Oriented Guarantees}
\label{subsec:results_certificate_audit}

The final part of this section audits the worst-case certificate in Eq.~\eqref{eq:energy_certificate_results} and characterizes its empirical behavior. Because $\Delta E_{\mathrm{bound}}(k)$ depends only on $\rho_k$ and $\|H_{\mathrm{tr}}\|_F$, it can be computed for every system and every scanned rank without reference energies. Reference data are needed only for audit. Certificate quantities are computed for the common PT subset ($N=350$ systems), while observed ground-state errors are available for a smaller evaluated subset because full diagonalization is only feasible for smaller instances; Appendix Table~\ref{tab:certificate_overview} summarizes this coverage.

\begin{figure*}[htbp]
    \centering
    \includegraphics[width=\textwidth]{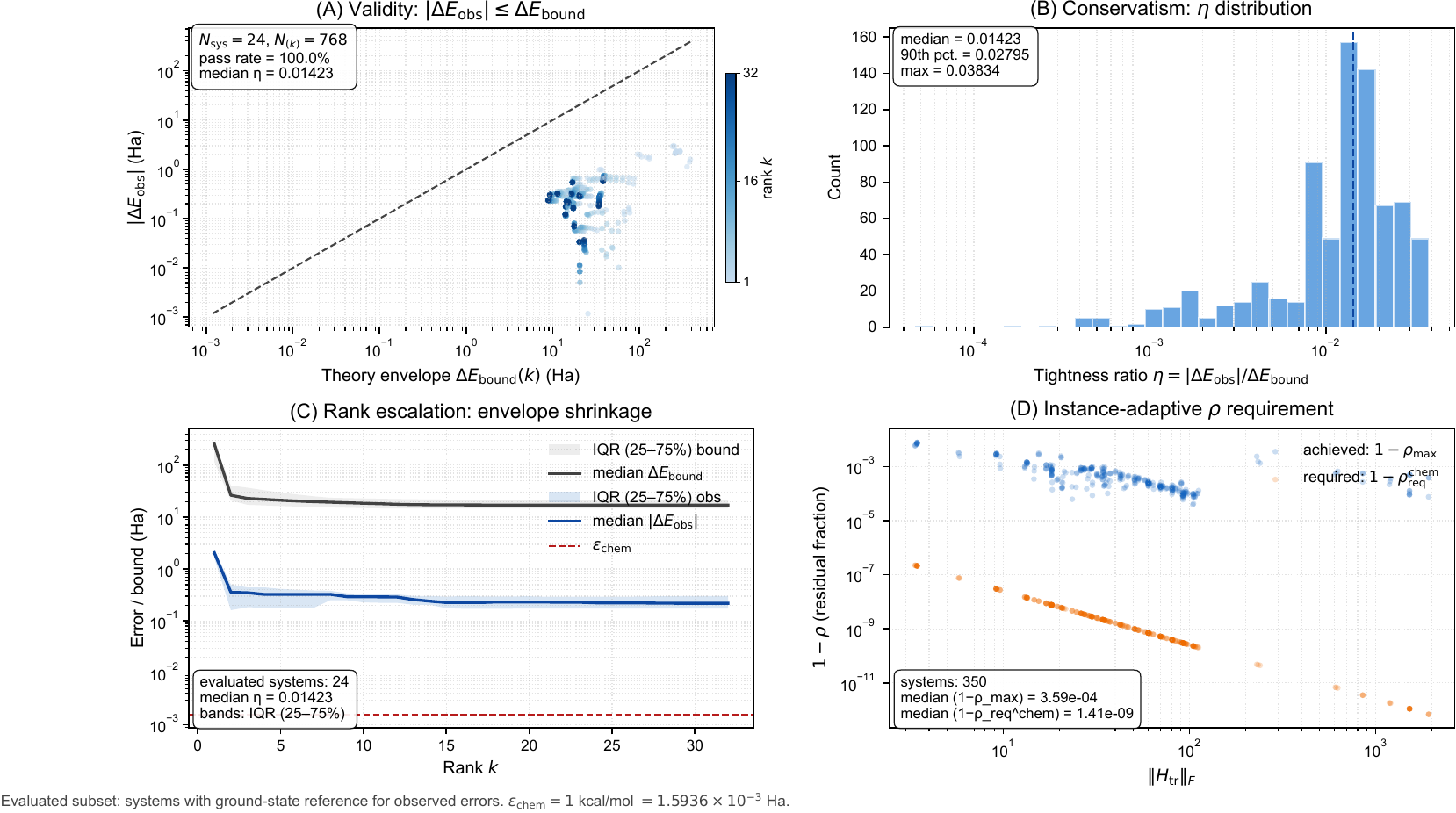}
    \caption{Certificate audit and deployment implications. (A) Observed ground-state errors versus the theoretical envelope demonstrate validity (all points lie below $y=x$). (B) Tightness ratio $\eta$ distribution quantifies conservatism. (C) Bound shrinkage with rank supports safe, auditable escalation. (D) Instance-adaptive chemical requirements show that worst-case chemical certification induces extremely strict residual targets that depend on $\|H_{\mathrm{tr}}\|_F$.}
    \label{fig:certificate_validity}
\end{figure*}

Figure~\ref{fig:certificate_validity} shows two empirical properties. First, the certificate is valid on every evaluated case: across $768$ $(\text{system},k)$ points from $24$ systems and $k\in[1,32]$, we observe no violations of $|\Delta E_{\mathrm{obs}}(k)|\le \Delta E_{\mathrm{bound}}(k)$. Second, the certificate is conservative. The tightness ratio $\eta=|\Delta E_{\mathrm{obs}}|/\Delta E_{\mathrm{bound}}$ has median $0.014$, a 90th percentile of $0.028$, and a maximum of $0.038$. The median ratio corresponds to observed ground-state errors that are about seventy times smaller than the worst-case envelope on typical evaluated cases. This is consistent with the derivation, which compounds a Frobenius-to-spectral relaxation with a worst-case eigenvalue perturbation bound.

These values indicate substantial conservatism. The certificate functions as a valid safety envelope, not as a sharp predictor of the realized error. Figure~\ref{fig:certificate_validity}(C) shows that the envelope shrinks monotonically with rank, so increasing $k$ does not weaken the bound. This monotonicity is consistent with the rank profiles in Figure~\ref{fig:rank_profiles}: the same concentration of capture at low rank also produces the first steep reductions in the auditable bound.

The same audit also shows that fixed global $\rho$ thresholds do not imply chemistry guarantees. Within the scanned budget $k\le 32$, $84.9\%$ of the common PT systems reach $\rho\ge 0.999$, yet none satisfy the worst-case chemical-accuracy requirement $\Delta E_{\mathrm{bound}}(k)\le \epsilon_{\mathrm{chem}}$ at that target, and the same remains true even at $\rho\ge 0.9999$ (Appendix Table~\ref{tab:certificate_overview}). This result reflects the scale dependence of the bound through $\rho_{\mathrm{required}}=1-(\epsilon_{\mathrm{chem}}/\|H_{\mathrm{tr}}\|_F)^2$. For a moderate $\|H_{\mathrm{tr}}\|_F=30$ Ha, one already needs $1-\rho_k\le (\epsilon_{\mathrm{chem}}/30)^2\approx 2.8\times 10^{-9}$, or $\rho_{\mathrm{required}}\approx 0.999999997$, which is far beyond standard global targets such as $0.999$ or $0.9999$. Figure~\ref{fig:certificate_validity}(D) shows that this required residual tightens rapidly as $\|H_{\mathrm{tr}}\|_F$ grows.

In practical use, the certificate provides an auditable decision rule: the rank or cut can be adjusted until the bound meets the required threshold, and systems that remain outside the threshold can be retained for the uncompressed reference treatment. To quantify the rank increase required under this rule, we ran a separate boundary experiment on a fixed, fully auditable set of twenty 12-qubit STO-3G cases. These cases already start from relatively high rank-1 traceless capture ($\rho_{1,\mathrm{tr}}\in[0.957,0.980]$), yet all $20/20$ require substantially larger ranks to certify chemical accuracy: the certification rank ranges from $25$ to $57$, with median $33$ and 90th percentile $51.1$, while the median time to certification is $66.5$ s and the 90th percentile is $213$ s. Relative to the PT mainline result $\mathrm{median}\,k^{\star}(0.999)=7$, this experiment quantifies the separation between high-fidelity approximation and worst-case chemical certification.

\begin{figure*}[htbp]
    \centering
    \includegraphics[width=\textwidth]{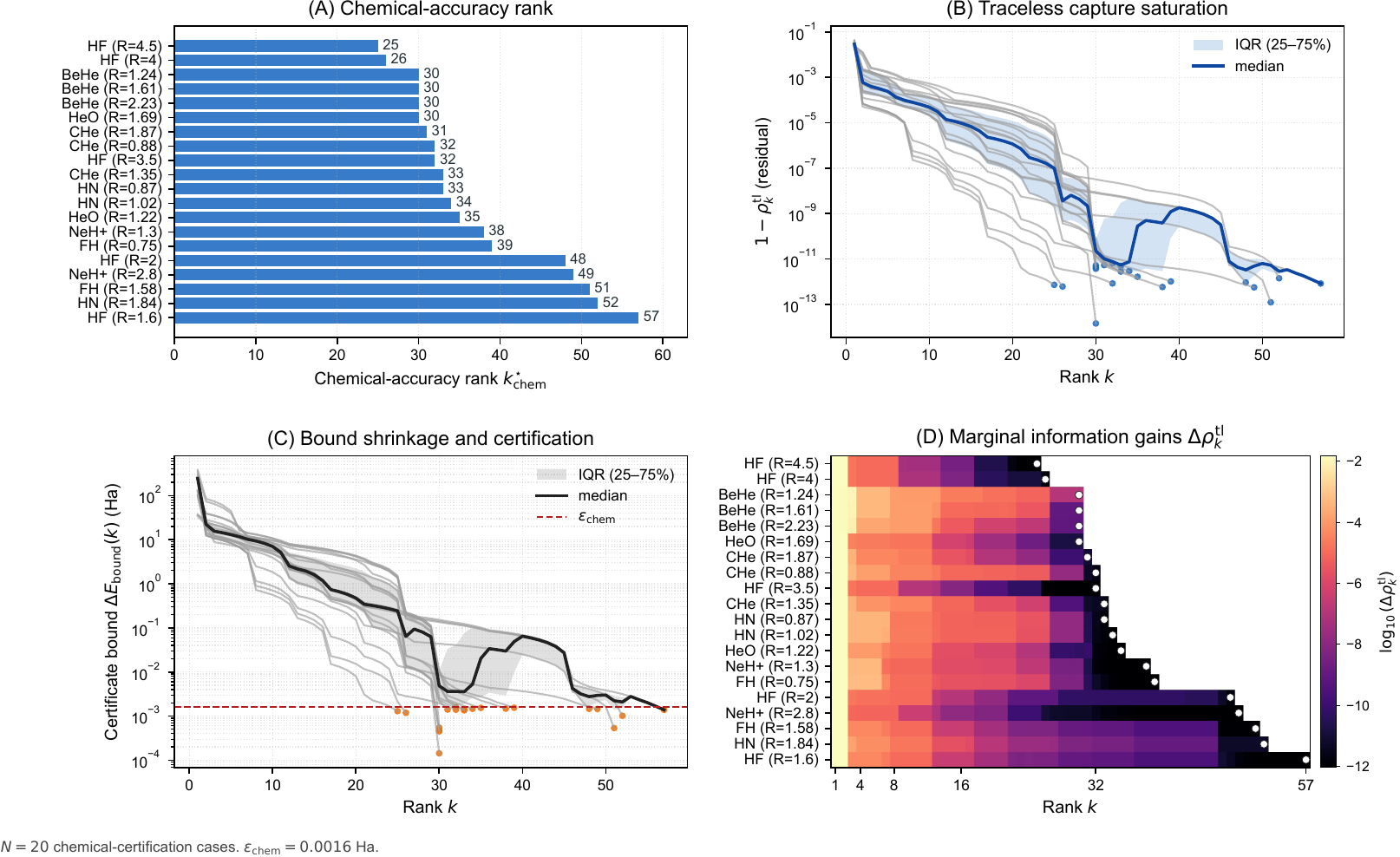}
    \caption{Chemistry-accuracy boundary test on 20 fixed cases. Panels report (A) the certification rank $k^{\star}_{\mathrm{chem}}$, (B) saturation of the traceless residual $1-\rho^{\mathrm{tl}}_k$, (C) shrinkage of the worst-case bound toward $\epsilon_{\mathrm{chem}}$, and (D) marginal information gains across rank, with markers at $k^{\star}_{\mathrm{chem}}$.}
    \label{fig:chem_cases}
\end{figure*}

Figure~\ref{fig:chem_cases} shows this separation directly. The traceless residual typically drops quickly in the early ranks, but the worst-case bound continues to decrease after the structural metric has already entered a high-capture regime. The boundary test is consistent with the main PT results and extends them to the certification regime: low ranks are sufficient to obtain large savings and high coefficient-space fidelity on many systems, whereas certification-level guarantees require larger ranks. Taken together, the screening, rank-profile, resource, and certificate results support using Qronecker as a controllable approximation primitive whose rank and cut are selected to meet explicit resource or guarantee targets, rather than a fixed global truncation rule.

\section{Discussion}
\label{sec:discussion}

\subsection{Interpretation and Positioning}
\label{subsec:discussion_interpretation}

This study shows that low-rank Kronecker structure in Pauli coefficient space is
not an isolated feature of a few favorable chemistry instances, but a recurring
property of mapped molecular Hamiltonians that can be converted into an auditable
computational decision process. The central contribution of Qronecker is therefore
not only a compression procedure, but a unified interface that couples screening,
rank selection, cut choice, resource accounting, and conservative certification
within the same coefficient-space representation.

This point distinguishes the present work from most existing reduction strategies.
Physical approximations such as frozen-core and active-space methods reduce model
size by changing the effective problem. Tensor and locality-based methods reduce
cost by exploiting structure at the integral, orbital, or operator-representation
level~\cite{weigend2009approximated,hohenstein2012tensor,motta2021low,mcclean2014exploiting}. Qronecker addresses a different stage of the computational chain: it operates
directly on the mapped qubit Hamiltonian and returns an instance-specific operating
point defined by the pair $(A\mid B,k)$ together with a computable worst-case energy
envelope. In that sense, the substantive advance is not solely lower cost, but the
conversion of structural compressibility into a deployable decision rule with an
explicit criterion for retaining the reference treatment when certification targets
are not met.

\subsection{Scientific Significance and Mechanistic Reading}
\label{subsec:discussion_significance}

The empirical trends reported in Section~\ref{sec:results} are consistent with the
coefficient-space formulation introduced in Section~\ref{sec:method}. After removal
of the identity offset, the traceless Hamiltonian is represented by the cut-aware
matrix $C(H_{\mathrm{tr}};A\mid B)$, and its singular spectrum directly determines
both the structural metric $\rho_k$ and the certificate in
Eq.~\eqref{eq:method_certificate}. The rapid rise of $\rho_k$ at low rank and the
fast decay of $\Delta \rho_k$ therefore indicate that, for many chemistry systems,
most coefficient-space weight is concentrated in a small number of dominant
cross-subsystem modes. This observation is physically plausible because mapped
electronic Hamiltonians originate from structured one- and two-body interactions
rather than from generic dense operators. Under favorable bipartitions, these
correlations remain sufficiently organized that a few singular directions capture
most of the traceless coefficient mass.

The heterogeneity observed across the boundary and stress cohorts is equally
informative. The lower screening scores in the stress cohort, the geometry-dependent
worst-case behavior across scan series, and the cut sensitivity reported in the
appendix together indicate that low-rank structure is prevalent but not universal.
This is consistent with the problem setup in Section~\ref{subsec:method_screen_rank_cut},
where cut and rank are treated as explicit decision variables rather than as fixed
global hyperparameters. In practical terms, the data support a conditional
statement: coefficient-space compression is broadly useful on mapped chemistry
Hamiltonians, but its effectiveness depends on the alignment between the molecular
instance, the subsystem partition, and the target accuracy level.

The resource trends follow from the same mechanism. Classical preprocessing gains
are largest because the dense baseline must materialize and factorize the full
coefficient matrix, whereas the decomposed route only targets the leading
coefficient-space modes. Once low-rank concentration is present, the difference in
storage and factorization cost grows rapidly with system size, which explains the
strong size dependence in Figures~\ref{fig:resource_benefits}
and~\ref{fig:rank_resource_tradeoff}. Circuit-side gains are more limited and decay
more quickly with rank because subsystem compilation overhead remains non-negligible
and the reduced subsystem circuits are also shaped by ansatz choice, compilation
heuristics, and subsystem imbalance. The disparity between classical and circuit-side
benefits is therefore expected: the coefficient-space approximation acts directly on
the classical preprocessing bottleneck, while quantum-circuit reductions are an
indirect consequence mediated by the chosen decomposition rank and cut.

The certificate results close the methodological loop. In the method section, the
certificate is derived from a Frobenius residual identity followed by a
Frobenius-to-spectral relaxation and a worst-case eigenvalue perturbation bound.
The empirical audit confirms the expected consequence of this derivation: the
certificate is valid across all evaluated records, but it is also substantially
conservative. This conservatism is not a defect in the logic of the bound; it is the
price of obtaining a state-independent guarantee available for every system from
coefficient-space quantities alone. The chemistry-boundary experiment clarifies the
practical implication. Because the required residual scales with
$\epsilon_{\mathrm{chem}}/\|H_{\mathrm{tr}}\|_F$, a fixed global target such as
$\rho=0.999$ or $0.9999$ cannot serve as a universal surrogate for chemistry-level
certification. The observed gap between high-fidelity truncation and certification
rank is therefore a direct consequence of the scale dependence of the certificate,
not an inconsistency between the structural and chemical analyses.

Taken together, these findings define the role of Qronecker in a broader quantum
chemistry setting. The method is best understood as a controllable operator
preprocessing layer. It identifies when mapped Hamiltonians lie in a favorable
low-rank regime, quantifies how much rank is needed to achieve a target structural
or certified accuracy level, and provides a conservative acceptance criterion when
chemistry-level guarantees are required. This is a different objective from
constructing a universally aggressive compression scheme, and the present data
support that more selective operational interpretation.

\subsection{Limitations}
\label{subsec:discussion_limitations}

Several limitations should be stated explicitly. First, the present empirical study
does not cover the full diversity of chemistry Hamiltonians. The boundary-scan and
performance-test cohorts are broad enough to establish recurring low-rank structure,
but they remain dominated by STO-3G systems and extend only to the system sizes that
can be processed and audited within the current computational budget. The fixed
twenty-case chemistry-boundary study provides a controlled certification benchmark,
but it does not exhaust the range of strongly correlated, large-basis, open-shell,
or symmetry-broken regimes that may be encountered in broader electronic-structure
applications.

The predominance of STO-3G should be interpreted carefully. It is not introduced to
favor unusually small or easy examples, but follows from the validation standard
adopted in this paper. The main claims are benchmarked against dense SVD wherever a
classical preprocessing baseline is required, and energy-oriented audit claims are
checked against exact diagonalization on the size-feasible subset. Both reference
procedures impose strict memory and wall-time ceilings. As a result, the number of
large-basis instances that can be carried through a like-for-like audit is
substantially smaller than the number that can be processed by Qronecker itself. The
restriction toward STO-3G in the fully auditable subsets is therefore an objective
tradeoff made in favor of the strongest available verification standard, not an
attempt to leave a gap in the difficulty of the tested instances.

Second, the method depends materially on the subsystem bipartition. The appendix
shows that both compressibility and realized resource factors vary across cuts, and
the current work evaluates only a limited set of balanced and fixed unbalanced
partitions. The study therefore demonstrates cut dependence clearly, but it does not
yet solve the global cut-selection problem. A poor cut can reduce or even remove the
storage advantage despite nontrivial structural compressibility, so cut optimization
remains an open problem for practical deployment.

Third, several operating thresholds used in the experiments are study-level design
choices rather than universal constants. Screening thresholds such as
$\rho_{1,\mathrm{tr}}\ge 0.85$, target values such as $\tau=0.999$ or $0.9999$, and
the chemistry-boundary optimization settings are reported explicitly, but their best
values may vary with dataset composition, basis choice, target observable, and the
intended deployment regime. The current results show that the overall decision
framework is robust to these choices, but they do not imply threshold invariance
across all chemistry families.

Fourth, the certificate is intentionally conservative. Its validity is an important
strength, but the same conservatism can force rank escalation well beyond the level
needed for typical observed energy errors. In practice, this means that the current
certificate is more suitable as a safety envelope and decision rule than as a
sharp predictor of realized error. Applications that require less conservative
certification may need tighter operator-norm estimates, state-aware bounds, or
hybrid audit procedures.

Fifth, the downstream quantum-resource study is based on compiled-circuit surrogates
and stratified subsets rather than on a complete end-to-end hardware benchmark. The
reported depth and two-qubit gate reductions are informative for relative scaling,
but they do not by themselves determine final wall-clock runtime, noise sensitivity,
or estimator variance on actual hardware. The present results therefore establish
that coefficient-space compression can influence downstream circuit structure, but
they do not yet quantify the full application-level impact in a noisy execution
environment.

\subsection{Future Directions}
\label{subsec:discussion_future}

The most immediate next step is to strengthen the adaptive decision layer. This
includes systematic cut optimization, joint cut-rank selection, and screening models
that use richer structural statistics than $\rho_{1,\mathrm{tr}}$ alone. Such
extensions would move the method from a fixed-policy benchmark toward a more general
adaptive decision procedure that automatically selects the most favorable compression
regime for each instance.

A second direction is to develop tighter certification tools. The present
Frobenius-based bound is attractive because it is general and inexpensive, but the
results indicate that chemistry-level deployment would benefit from less conservative
audits. Promising directions include operator-norm surrogates, variationally
informed residual bounds, observable-specific certificates, and multi-stage schemes
in which a coarse certificate governs screening and a tighter local audit is applied
only to systems near the target threshold.

A third direction is broader empirical coverage. The current study should be
extended to larger basis sets, more strongly correlated species, longer reaction
coordinates, excited-state settings, and other fermion-to-qubit mappings. The same
coefficient-space framework could also be tested on lattice-model Hamiltonians and
other structured many-body operators, where cut-aware low-rank decompositions may
serve a similar role as preprocessing and certification primitives.

Finally, the most consequential application direction is end-to-end integration into
hybrid quantum algorithms for chemistry. In SQD, variational
eigensolvers~\cite{peruzzo2014variational}, related subspace methods,
and fragment-based quantum chemistry, Qronecker could be used to determine
in advance which instances merit compression, which cut/rank pair is compatible with
the requested accuracy, and when the reference treatment should be retained. In
longer-term molecular screening or drug-discovery studies, this type of auditable
operator preprocessing could support triage across large numbers of candidate systems
by allocating tighter resources only where the certificate indicates that compression
remains safe. The present study does not complete that integration, but it
establishes the structural and methodological basis for it.

\section{Method}
\label{sec:method}

The empirical study is organized as six coordinated experiments: boundary screening
of low-rank structure, rank-profile analysis through sparse rank-$k$ decomposition,
cut-sensitivity analysis, circuit-resource evaluation, certificate audit, and
chemistry-accuracy boundary testing. These components share a common
coefficient-space formulation and are applied to the same family of molecular
Hamiltonians. Taken together, they quantify the full chain from structural screening
to rank budgeting, resource measurement, and certificate-governed deployment. For
each molecular geometry,
basis, charge, and spin specification,
we obtain one- and two-electron integrals with PySCF~\cite{sun2020recent,sun2018pyscf}, assemble the electronic
Hamiltonian, and map it to qubit form through OpenFermion~\cite{mcclean2020openfermion} and the Jordan--Wigner
transform~\cite{jordan1928paulische}. All screening, rank selection, resource accounting, and certification
steps are then performed in Pauli coefficient space, so that the same quantities
support the structural, computational, and chemistry-oriented analyses.

\subsection{Problem Setup and Notation}
\label{subsec:method_problem_setup}

Let the mapped $n$-qubit Hamiltonian be written as a Pauli sum
\begin{equation}
H = \sum_{p \in \mathcal{P}_n} c_p P_p,
\qquad
\mathcal{P}_n = \{I,X,Y,Z\}^{\otimes n},
\end{equation}
with real coefficients for the chemistry instances considered here. We separate the
identity offset and work with the traceless operator
\begin{equation}
c_0 \equiv c_{I^{\otimes n}},
\qquad
H_{\mathrm{tr}} \equiv H - c_0 I.
\end{equation}
Unless noted otherwise, every compressibility and certificate quantity reported in
this paper is computed on $H_{\mathrm{tr}}$ rather than on the full operator $H$.

Fix a bipartition $A \mid B$ with $n_A + n_B = n$, where $n_A$ is either prescribed
or chosen as the balanced split $n_A = \lfloor n/2 \rfloor$ by default. Any Pauli
string factorizes uniquely as $P_p = P_a \otimes P_b$, with
$P_a \in \{I,X,Y,Z\}^{\otimes n_A}$ and
$P_b \in \{I,X,Y,Z\}^{\otimes n_B}$. We reshape the traceless coefficients into the
cut-dependent matrix
\begin{equation}
C(H_{\mathrm{tr}};A \mid B) \in \mathbb{R}^{4^{n_A} \times 4^{n_B}},
\qquad
C_{a,b} \equiv c_{(a,b)} .
\end{equation}
Under the unnormalized Pauli basis,
$\mathrm{Tr}(P_p P_q) = 2^n \delta_{pq}$, so the operator and coefficient norms are
related exactly by
\begin{equation}
\|H_{\mathrm{tr}}\|_F^2 = 2^n \|C\|_F^2 .
\end{equation}

Let
\begin{equation}
C = \sum_{i \ge 1} \sigma_i u_i v_i^{\top},
\qquad
\sigma_1 \ge \sigma_2 \ge \cdots \ge 0,
\end{equation}
be the singular value decomposition of the cut-dependent coefficient matrix. The
rank-$k$ truncation
\begin{equation}
C_k \equiv \sum_{i=1}^{k} \sigma_i u_i v_i^{\top}
\end{equation}
induces the rank-$k$ Kronecker approximation
\begin{equation}
\widetilde{H}_{k,\mathrm{tr}}
\equiv
\sum_{i=1}^{k} \sigma_i \big(A_i \otimes B_i\big),
\qquad
\widetilde{H}_k \equiv c_0 I + \widetilde{H}_{k,\mathrm{tr}},
\end{equation}
where $A_i$ and $B_i$ are subsystem Pauli-sum operators obtained by reshaping the
singular vectors under a fixed basis ordering.

We quantify instance-wise compressibility through the cumulative captured energy
\begin{equation}
\rho_k
\equiv
\frac{\|C_k\|_F^2}{\|C\|_F^2}
=
\frac{\sum_{i=1}^{k} \sigma_i^2}{\sum_{i \ge 1} \sigma_i^2}
\in [0,1],
\qquad
\rho_{1,\mathrm{tr}} \equiv \rho_1(H_{\mathrm{tr}};A \mid B),
\end{equation}
the marginal gain
\begin{equation}
\Delta \rho_k \equiv \rho_k - \rho_{k-1},
\qquad
\rho_0 \equiv 0,
\end{equation}
and the first-hit rank for a target $\tau \in (0,1)$,
\begin{equation}
k^{\star}(\tau) \equiv \min \{ k : \rho_k \ge \tau \}.
\end{equation}
When a target is not reached within the scanned budget $k_{\max}$, the instance is
treated as right-censored at $k_{\max}$ rather than being assigned an extrapolated
rank.

\subsection{Coefficient-Space Qronecker Primitive}
\label{subsec:method_qronecker_primitive}

The central computational primitive is to operate on the sparse Pauli coefficient
matrix without materializing either the dense $2^n \times 2^n$ operator or the dense
$4^{n_A} \times 4^{n_B}$ reshape. After dropping coefficients smaller than a fixed
tolerance ($10^{-12}$ in the main experiments), we store the nonzero entries as
sparse triples
\begin{equation}
\mathcal{E} = \{(a_j,b_j,c_j)\}_{j=1}^{m},
\end{equation}
where $m$ is the number of nonzero Pauli pairs under the selected cut. This supports
implicit matrix--vector products
\begin{equation}
(Cx)_a = \sum_{j=1}^{m} c_j x_{b_j} \mathbf{1}[a_j=a],
\qquad
(C^{\top} z)_b = \sum_{j=1}^{m} c_j z_{a_j} \mathbf{1}[b_j=b],
\end{equation}
with cost linear in $m$.

Rank-1 screening uses sparse power iteration on $CC^{\top}$ and $C^{\top}C$ to obtain
the leading singular triplet. In the main boundary analysis, the screening solver uses
a balanced cut, coefficient tolerance $10^{-12}$, a maximum of $200$ power-iteration
steps, convergence tolerance $10^{-8}$, and deterministic initialization with seed
$0$. For rank-$k$ decomposition, we iteratively extract leading
components and deflate the residual,
\begin{equation}
C^{(0)} = C,
\qquad
(\sigma_r,u_r,v_r) = \mathrm{TopSVD}(C^{(r-1)}),
\qquad
C^{(r)} = C^{(r-1)} - \sigma_r u_r v_r^{\top},
\end{equation}
until either the target capture threshold is reached or the rank budget is exhausted.
The main sparse solver uses at most $300$ power-iteration steps, convergence tolerance
$10^{-10}$, and deterministic initialization with seed $0$; stricter validation runs
may raise the iteration limit when exact dense recovery is checked on small systems.

This primitive returns four quantities for every analyzed instance: the selected cut,
the full $\rho_k$ curve up to the requested rank, the subsystem factors
$\{A_r,B_r\}_{r \le k}$, and the residual statistics needed for later resource and
certificate calculations. The same representation is reused throughout the study,
which avoids re-deriving separate notions of approximation quality for screening,
benchmarking, and chemistry certification.

\subsection{Screening, Rank Selection, and Cut Policy}
\label{subsec:method_screen_rank_cut}

We use the traceless rank-1 statistic $\rho_{1,\mathrm{tr}}$ as the primary screening
variable. The identity contribution can dominate the total Frobenius norm of a qubit
Hamiltonian and artificially inflate a total-space rank-1 metric, whereas the later
certificate depends on the residual of $H_{\mathrm{tr}}$. Using the traceless matrix
therefore aligns the screening step with the quantity that controls the worst-case
energy envelope.

In the preprocessing and resource experiments, the screening rule is operationalized
through two thresholds. Systems with $\rho_{1,\mathrm{tr}} \ge 0.85$ are included in
the primary benchmark set; systems with $0.70 \le \rho_{1,\mathrm{tr}} < 0.85$ form a
secondary lower-compressibility set that can still be analyzed separately; systems
below $0.70$ are excluded from the main speed and resource benchmark. These
thresholds are workload-control rules rather than
theoretical guarantees, and they are reported explicitly so that the supported regime
is transparent.

Rank is selected by the smallest $k$ that reaches a target capture level
$k^{\star}(\tau)$. The main performance study scans all systems to $k_{\max}=32$
and evaluates operating points such as $\tau=0.999$ and $\tau=0.9995$, while the
integrated validation study uses $\tau=0.9999$ on the common subset and
$\tau=0.999$ on the stress subset. In addition to target-driven ranks, we also report
the full $\rho_k$ curve so that fixed-rank analyses can be carried out directly.

The cut is treated as an explicit decision variable. The balanced split is the default
because it controls factor dimensions symmetrically, but all quantities above depend
on the cut through $C(H_{\mathrm{tr}};A \mid B)$. We therefore run a dedicated
unbalanced-cut study in which fixed left-part sizes $n_A \in \{2,4,5\}$ are compared
under otherwise identical settings. This ablation makes it possible to separate
intrinsic low-rank structure from purely dimensional effects in storage, timing, and
compiled-circuit proxies.

\subsection{Certificate Derivation and Audit Metrics}
\label{subsec:method_certificate}

The coefficient-space residual yields an exact Frobenius residual identity~\cite{eckart1936approximation,johnson1963theorem},
\begin{equation}
\|C - C_k\|_F
=
\|C\|_F \sqrt{1-\rho_k}
\quad \Longrightarrow \quad
\|H_{\mathrm{tr}} - \widetilde{H}_{k,\mathrm{tr}}\|_F
=
\|H_{\mathrm{tr}}\|_F \sqrt{1-\rho_k}.
\end{equation}
For any normalized state $|\psi\rangle$,
\begin{equation}
\left| \langle \psi | (H-\widetilde{H}_k) | \psi \rangle \right|
\le
\|H-\widetilde{H}_k\|_2
\le
\|H_{\mathrm{tr}}-\widetilde{H}_{k,\mathrm{tr}}\|_F ,
\end{equation}
which gives the conservative certificate
\begin{equation}
\Delta E_{\mathrm{bound}}(k)
\le
\|H_{\mathrm{tr}}\|_F \sqrt{1-\rho_k}.
\label{eq:method_certificate}
\end{equation}
Equation~\eqref{eq:method_certificate} is the only bound used for the global
certificate claims in the paper. It is available for every system once $\rho_k$ and
$\|H_{\mathrm{tr}}\|_F$ are known, regardless of whether an exact reference energy is
computationally accessible.

When reference energies are available, we audit the conservatism of the certificate
through tightness ratios. For ground-state validation we use
\begin{equation}
\eta_{\mathrm{gs}}(k)
\equiv
\frac{|E_0(H)-E_0(\widetilde{H}_k)|}{\Delta E_{\mathrm{bound}}(k)},
\end{equation}
and for state-ensemble checks we use
\begin{equation}
\mathrm{Err}_{\mathrm{rms}}(k)
\equiv
\sqrt{\frac{1}{N}\sum_{i=1}^{N}
\left|
\langle \psi_i | (H-\widetilde{H}_k) | \psi_i \rangle
\right|^2},
\qquad
\eta_{\mathrm{rms}}(k)
\equiv
\frac{\mathrm{Err}_{\mathrm{rms}}(k)}{\Delta E_{\mathrm{bound}}(k)}.
\end{equation}
A certificate evaluation is counted as a pass when the relevant ratio is at most one.
For small instances that remain amenable to dense evaluation, we additionally estimate
$\|\Delta H_k\|_2$ by power iteration and retain the tighter operator-norm comparison
as an auxiliary audit, but the
paper's safety claim is based on Eq.~\eqref{eq:method_certificate}.

For chemistry-oriented discussion we use the standard threshold
\begin{equation}
\epsilon_{\mathrm{chem}} = 1 \ \mathrm{kcal/mol}
\approx
1.5936 \times 10^{-3} \ \mathrm{Ha},
\end{equation}
which induces the instance-wise requirement
\begin{equation}
\rho_{\mathrm{required}}
\equiv
1-\left(\frac{\epsilon_{\mathrm{chem}}}{\|H_{\mathrm{tr}}\|_F}\right)^2 .
\end{equation}
The chemistry-boundary calculations use the rounded numerical value
$1.6 \times 10^{-3}$ Ha; this differs only in the fourth significant digit and does
not change the qualitative behavior of the certification boundary.

\subsection{Resource Models and Measurement Protocols}
\label{subsec:method_resource_protocol}

\paragraph{Classical preprocessing experiment.}
The primary classical baseline is dense singular-value decomposition of the full
coefficient matrix $C$. Dense SVD is attempted only when the matrix is small enough to
materialize safely; in the primary preprocessing benchmark this is enforced by the limits
$\max(4^{n_A},4^{n_B}) \le 5500$ and at most $3.2 \times 10^{7}$ matrix elements.
Timing is summarized by the median over repeated runs after warmup (five repeats and
one warmup in the main preprocessing benchmark; three repeats and one warmup in the
integrated validation study). We report
\begin{equation}
\mathrm{Speedup}
\equiv
\frac{t_{\mathrm{dense}}}{t_{\mathrm{decomp}}},
\qquad
\mathrm{Memory\ Ratio}
\equiv
\frac{\mathrm{bytes}_{\mathrm{dense}}}{\mathrm{bytes}_{\mathrm{factors}}},
\end{equation}
where $\mathrm{bytes}_{\mathrm{factors}}$ stores the retained singular weights and the
subsystem Pauli factors rather than the dense matrix.

\paragraph{Circuit-resource experiment.}
Circuit evaluations are carried out on a stratified subset drawn from the common and
stress pools, with target sample size $48$ and stress quota $12$, subject to the
additional filters $\rho_{1,\mathrm{tr}} \ge 0.85$ and an upper bound on qubit count
for circuit construction. We first construct a chemistry-motivated UCCSD ansatz from
the molecular data. If that construction is unavailable, it is replaced by a
hardware-efficient parameterized circuit according to a fixed prespecified fallback
hierarchy while keeping the compilation settings identical across full and cut
circuits. Compilation is carried out at optimization level $1$ under a linear
nearest-neighbor coupling topology with gate set
$\{R_z,\sqrt{X},X,\mathrm{CX},\mathrm{SWAP}\}$ and fixed compiler seed $11$.

For a full $n$-qubit circuit and its cut proxy circuits on subsystems $A$ and $B$, we
measure the two-qubit depth and gate counts after transpilation. The main reported
proxy metrics are the parallel cut depth
\begin{equation}
\mathrm{depth}_{\mathrm{cut,par}}
\equiv
\max(\mathrm{depth}_{A}, \mathrm{depth}_{B}),
\end{equation}
the summed two-qubit proxy count
\begin{equation}
\#2\mathrm{Q}_{\mathrm{cut}} \equiv \#2\mathrm{Q}_{A} + \#2\mathrm{Q}_{B},
\end{equation}
and the corresponding gains
\begin{equation}
\mathrm{Depth\ Gain}
\equiv
\frac{\mathrm{depth}_{\mathrm{full}}}{\mathrm{depth}_{\mathrm{cut,par}}},
\qquad
\mathrm{2Q\ Gain}
\equiv
\frac{\#2\mathrm{Q}_{\mathrm{full}}}{\#2\mathrm{Q}_{\mathrm{cut}}}.
\end{equation}
Sequential-depth and SWAP-sensitive variants are retained as supplementary diagnostics
but are not the primary quantities used in the main analysis.

\paragraph{Noiseless reconstruction and bound-alignment experiment.}
Small-system reconstruction is evaluated separately on instances with at most
$12$ qubits, where measurement simulation remains tractable. These checks select the
minimal rank that reaches the requested $\rho$ target, use $10^6$ shots, and compare
reconstructed energies against full-circuit reference values under absolute and
relative tolerances $10^{-2}$ Ha and $5 \times 10^{-3}$, respectively. In the
cut-sensitivity study, bound-alignment experiments use an approximate global
2-design ensemble of depth $6$, a target grid
$\tau \in \{0.85,0.90,0.95,0.98,0.99\}$, and $40$ power-iteration steps for the
optional operator-norm audit on residuals that remain amenable to dense evaluation.

\subsection{Datasets, Cohorts, and Experimental Design}
\label{subsec:method_datasets}

The empirical study uses two main cohorts. The boundary-scan cohort (BS) contains
$765$ molecular systems and serves as the broad applicability pool used to
characterize how often rank-1 structure appears and how it changes along recoverable
geometry scans. The performance-test cohort (PT) contains $400$ systems and is
derived from BS by a family-controlled selection procedure. Starting from $566$
candidate systems with sufficiently high rank-1 traceless capture, we construct a
common subset of $350$ systems and a stress subset of $50$ systems under explicit
family caps to avoid over-representing near-duplicate scan points: the common subset
requires
$\rho_{1,\mathrm{tr}} \ge 0.90$, at least $10$ qubits, and at most four members per
geometry family; the stress subset requires $\rho_{1,\mathrm{tr}} \ge 0.92$, at least
$20$ qubits, and at most one member per family. Stress cases are removed from the
common pool before the final $350+50$ selection.

\begin{figure}[htbp]
    \centering
    \includegraphics[width=\textwidth]{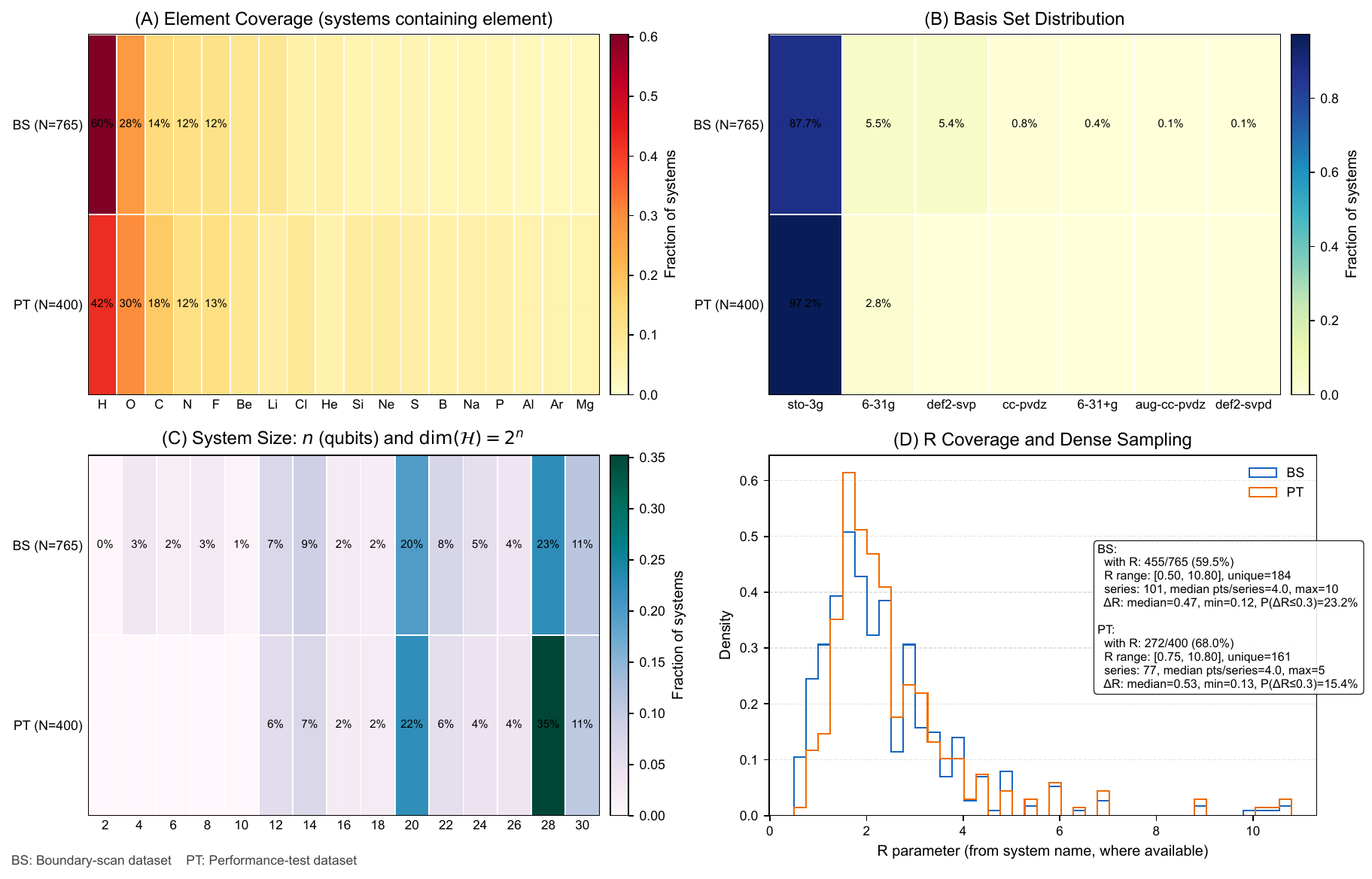}
    \caption{Dataset composition and cohort design for the boundary-scan and
    performance-test studies. The boundary-scan cohort provides the broad
    screening pool, while the performance-test cohort is a family-controlled
    high-compressibility subset used for rank, resource, and certificate
    experiments. The figure summarizes basis composition, qubit-count
    distribution, elemental coverage, and recoverable interatomic-distance scan
    coverage.}
    \label{fig:method_dataset}
\end{figure}

Figure~\ref{fig:method_dataset} summarizes the two-cohort design. The BS cohort is
used to estimate screening prevalence and scan robustness. The PT cohort is used for
rank profiles, classical speed and memory measurements, circuit-resource evaluation,
and certificate calculations. Within PT, the common subset carries the primary
high-compressibility study, while the stress subset covers the larger and more
computationally demanding systems.

The predominance of STO-3G in the fully auditable subsets is a consequence of the
validation protocol rather than of case selection for convenience. The study requires
direct comparison against dense SVD wherever classical preprocessing claims are made,
and exact diagonalization or equivalent dense reference energies wherever energy-audit
claims are reported. These baselines impose explicit resource ceilings: dense SVD is
attempted only under the matrix-size limits in
Section~\ref{subsec:method_resource_protocol}, and noiseless reference-energy
validation is restricted to systems with at most $12$ qubits. Larger-basis systems are
therefore included where feasible, but the strictest audit tracks necessarily contract
toward smaller mapped dimensions and are correspondingly dominated by STO-3G cases.

The experimental program is organized as follows:
\begin{enumerate}
    \item The boundary-screening experiment uses BS to estimate the prevalence of
    rank-1 structure and its robustness along geometry scans.
    \item The rank-profile experiment uses the full PT cohort, scans ranks up to
    $k_{\max}=32$, and measures the ranks required to reach prespecified capture
    targets.
    \item The resource experiment uses PT to compare coefficient-space decomposition
    against dense preprocessing and to measure the effect of decomposition on
    compiled-circuit proxies.
    \item The cut-sensitivity experiment repeats the structural and resource analyses
    under balanced and unbalanced bipartitions.
    \item The certificate-audit experiment computes bounds for all PT systems and
    compares them with observed errors on the size-feasible subset where dense
    diagonalization remains tractable.
    \item The integrated validation study uses the $350$-system common subset for the
    primary rank and resource analyses and the $50$-system stress subset for the
    higher-cost validation track, with default targets $\tau=0.9999$ and
    $\tau=0.999$, respectively.
\end{enumerate}
No instance is removed after observing its downstream performance; the only
post-construction exclusions arise from the prespecified inclusion criteria,
qubit-count limits, and dense-matrix size limits.

\subsection{Chemistry-Ceiling Boundary Test}
\label{subsec:method_chem_boundary}

To quantify the difference between high-fidelity approximation and worst-case
chemistry certification, we run a separate rank-escalation experiment in full Pauli
coefficient space. The evaluation set is defined prospectively by restricting the
high-compressibility subset to eligible systems with at most $12$ qubits and then
taking the first $20$ cases under the predetermined cohort ordering. In the analyzed
run, this selection yields a fully auditable set of twenty $12$-qubit STO-3G cases. No
instance-specific retuning or post hoc replacement is performed.

Unlike the sparse-support SVD primitive, the chemistry boundary test optimizes factors
in the full subsystem Pauli spaces. For rank $R$, we parameterize
\begin{equation}
\widehat{C}_R
\equiv
\sum_{r=1}^{R} \alpha_r \big(a_r \otimes b_r\big),
\qquad
a_r \in \mathbb{R}^{4^{n_A}},
\qquad
b_r \in \mathbb{R}^{4^{n_B}},
\end{equation}
and define the full-space residual $D_R \equiv C-\widehat{C}_R$. This formulation is
not restricted to the sparse support of the original Hamiltonian and can therefore
introduce compensating Pauli pairs if they reduce the global residual.

Each rank stage is optimized with the mixed objective
\begin{equation}
\mathcal{L}
=
\lambda_f \|D_R\|_F^2
+ \lambda_{\mathrm{spec}} \frac{1}{2} \sigma_{\max}(D_R)^2
+ \lambda_{\mathrm{tr}} \|\theta-\theta_0\|_2^2,
\label{eq:method_mixed_objective}
\end{equation}
where $\theta$ collects all current factor parameters and $\theta_0$ is the
stage-initial reference point. The first rank is initialized by an SVD-based starting
point, and each new rank is initialized from the leading singular direction of the
current residual.
The three terms play distinct roles in the chemistry-certification setting. The
Frobenius term is the primary fitting objective because it directly controls the
global coefficient-space residual and therefore the auditable upper bound
$\Delta E_{\mathrm{upper}}^{\mathrm{chem}}(R)=\sqrt{2^n}\|D_R\|_F$. The spectral
term suppresses residual concentration along a single dominant singular direction.
This is useful because chemistry certification is a worst-case requirement: two
solutions with similar Frobenius residuals can differ materially in how strongly the
remaining error is concentrated in the most adversarial direction. The trust-region
term stabilizes rank-by-rank continuation by discouraging unnecessarily large
parameter excursions away from the stage-initial reference point, which reduces
optimization drift after a new factor is introduced.

This mixed objective is used because a purely Frobenius-driven fit was found to be a
weaker proxy for certification-oriented convergence. In particular, optimizing only
$\|D_R\|_F^2$ can reduce average residual mass while leaving a relatively sharp
leading residual mode, which slows the shrinkage of the worst-case envelope. Adding a
moderate spectral penalty promotes more uniform residual suppression without changing
the final auditable stopping rule, and the weak trust-region regularization makes that
multi-stage optimization numerically more stable.

Optimization uses Adam with plateau-triggered learning-rate decay. For the fixed
$20$-case experiment, the run configuration is
\begin{equation}
\lambda_f = 1.0,
\qquad
\lambda_{\mathrm{spec}} = 0.05,
\qquad
\lambda_{\mathrm{tr}} = 10^{-4},
\end{equation}
with at most $300$ optimization steps per rank, a maximum examined rank of $512$,
spectral-term updates every $5$ iterations, and deterministic initialization with seed
$0$.
These weights reflect the intended priority of the three terms. Setting
$\lambda_f=1$ makes the certified Frobenius residual the dominant quantity
throughout. The smaller choice $\lambda_{\mathrm{spec}}=0.05$ is used to bias the
fit away from residuals with an excessively large leading singular direction while
avoiding a regime in which the spectral penalty overwhelms overall residual
reduction. The much smaller value $\lambda_{\mathrm{tr}}=10^{-4}$ keeps the
regularization term active only as a local stabilizer, so that it constrains
stage-to-stage motion without materially changing the target approximation itself.
Accordingly, the mixed objective acts as an optimization surrogate for efficient
certification-oriented fitting, whereas the actual chemistry pass/fail decision
continues to be determined solely by the auditable bound
$\Delta E_{\mathrm{upper}}^{\mathrm{chem}}(R)$ below.

At every rank stage, the Frobenius residual in coefficient space yields the auditable
upper bound
\begin{equation}
\Delta E_{\mathrm{upper}}^{\mathrm{chem}}(R)
\equiv
\sqrt{2^n} \, \|D_R\|_F ,
\end{equation}
which matches the worst-case energy envelope for the reconstructed operator. The run
stops at the first rank for which
\begin{equation}
\Delta E_{\mathrm{upper}}^{\mathrm{chem}}(R) \le \epsilon_{\mathrm{chem}} .
\end{equation}
For each system we record the certification rank, the time to first certification,
the bound value at the hit rank, the corresponding $\rho^{\mathrm{tl}}_R$, and the
full per-rank trace. This experiment quantifies the
rank escalation required when the deployment goal is a worst-case chemistry
guarantee instead of a fixed global $\rho$ target.

\subsection{Statistical Reporting and Reproducibility}
\label{subsec:method_repro}

All reported summaries are descriptive and auditable. Distributional plots use
empirical cumulative distributions, medians, percentile bands, or per-system traces as
appropriate to the experiment. Timing results are aggregated by medians across repeated
measurements after warmup. Target reachability is reported as exact pass rates over the
eligible denominator, and unreached systems remain explicit in the data rather than
being imputed.

All experiments were executed as single-node CPU jobs on the Ascend cluster
(``nextgen'' partition). The partition comprises Dell PowerEdge XE8645 nodes
equipped with two AMD EPYC 7643 (Milan) processors at $2.3$~GHz, providing $88$
usable CPU cores and $921$~GB usable memory per node, and Dell PowerEdge R7545
nodes equipped with two AMD EPYC 7H12 processors at $2.60$~GHz, providing $120$
usable CPU cores and $472$~GB usable memory per node. No GPU resources were used in
the present study. Thread-parallel linear-algebra libraries were restricted to the
requested core count to avoid oversubscription. The main run
configurations were $12$ CPU threads for boundary screening, $18$ CPU threads for
the sparse preprocessing benchmark, $6$ CPU threads for circuit-resource evaluation,
$12$ CPU threads for noiseless reconstruction, $20$ CPU threads with an explicit
$96$~GB memory reservation for the integrated validation and cut-sensitivity
pipelines, and $32$ CPU threads with $96$~GB for the chemistry-boundary experiment.

Randomized components were controlled by fixed pseudo-random seeds. The main
choices were seed $0$ for decomposition routines and mixed-objective optimization,
seed $20260221$ for stochastic subsampling in the integrated validation study, and
seed $11$ for circuit compilation. Intermediate quantities required for audit and
figure generation were
retained in structured experiment records, including boundary summaries, rank maps,
decomposition summaries, circuit-evaluation summaries, certificate-audit tables, and
full per-rank traces for the chemistry-boundary study. This record structure makes
each reported figure and summary statistic reproducible from the archived experiment
outputs rather than from transient interactive state.

\clearpage
\FloatBarrier

\bibliographystyle{unsrt}
\bibliography{references}

\clearpage
\FloatBarrier
\appendix
\clearpage
\onecolumn

\section{Supplementary Figures}
\label{sec:appendix_figures}

This appendix collects supporting figures and tables referenced from Section~\ref{sec:results}. The two figures below are moved out of the main Results so that the main section contains no more than six inserted figures and tables while the per-system heterogeneity and cut-sensitivity evidence remain directly auditable.

\begin{figure}[htbp]
    \centering
    \includegraphics[width=\textwidth]{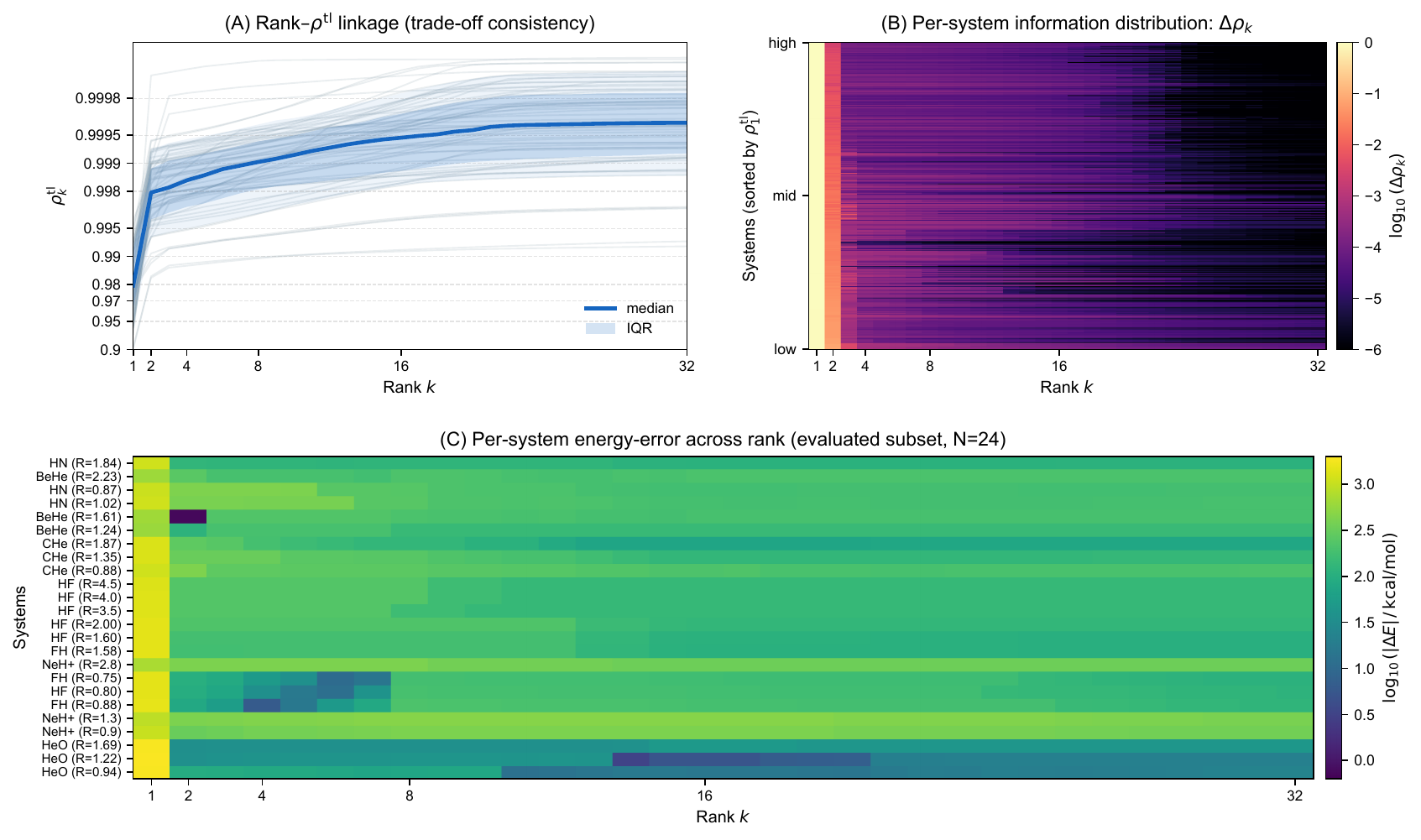}
    \caption{Per-system evidence for information concentration and heterogeneity on the performance-test cohort. (A) Representative per-system $\rho_k$ trajectories with median and IQR bands support a consistent rank--$\rho$ trade-off axis. (B) Heatmap of $\log_{10}(\Delta\rho_k)$ across systems and ranks reveals front-loaded contributions with long-tail corrections and system-to-system variability. (C) Heatmap of $\log_{10}(|\Delta E|)$ on the evaluated subset illustrates heterogeneous energy-error convergence patterns across rank.}
    \label{fig:rank_profiles_addon}
\end{figure}

\begin{figure}[htbp]
    \centering
    \includegraphics[width=\textwidth]{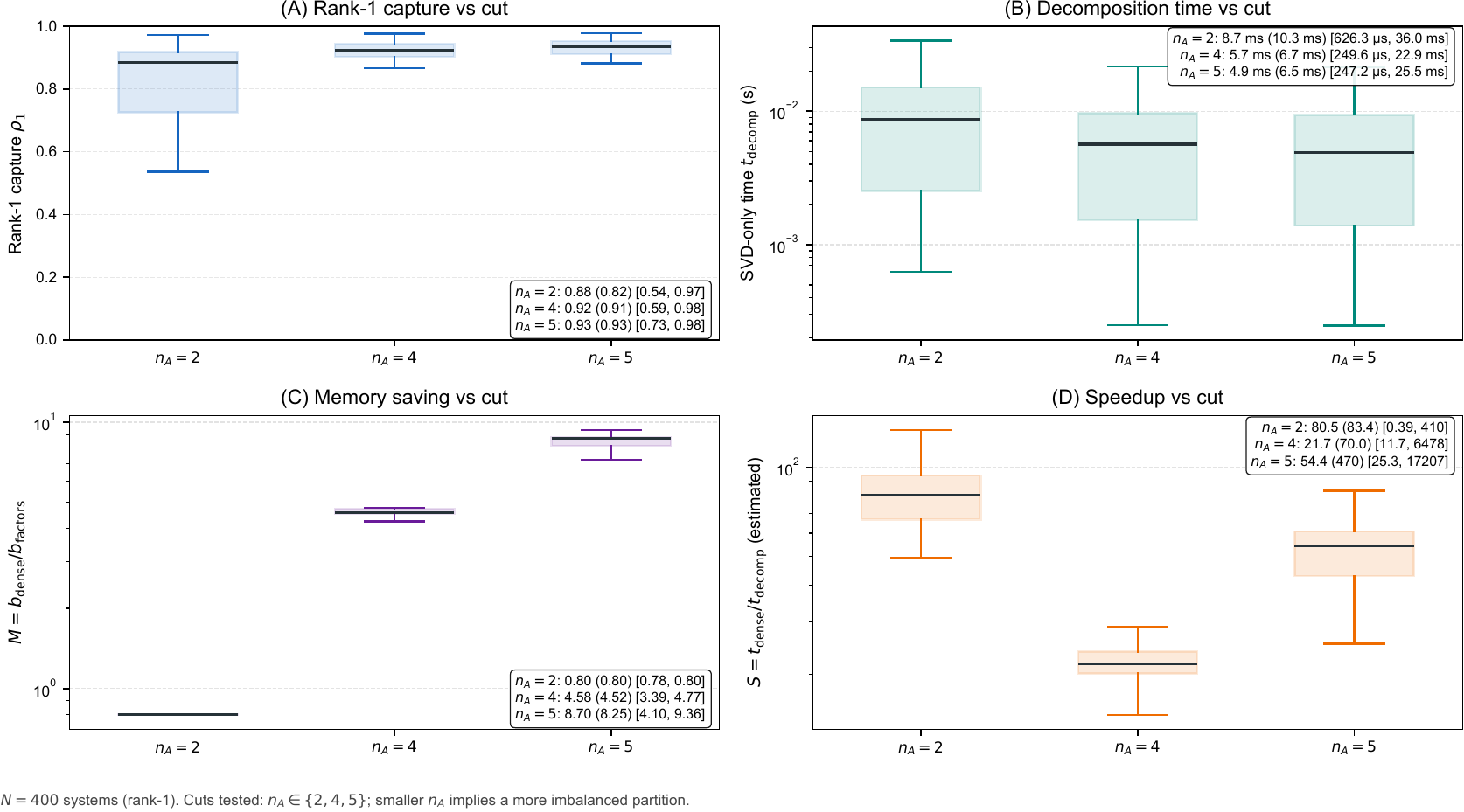}
    \caption{Cut sensitivity at rank~1 on the performance-test cohort. Across three tested partitions ($n_A\in\{2,4,5\}$), the rank-1 capture indicator $\rho_1$, decomposition time, memory ratio, and implied speedup vary substantially, demonstrating that partition choice is a concrete factor in both structural screening and realized resource outcomes.}
    \label{fig:cut_sensitivity}
\end{figure}

\FloatBarrier

\clearpage
\section{Supplementary Summary Tables}
\label{sec:appendix_summary_tables}

The summary tables used to define cohort coverage, representative operating points, and certificate-audit totals are collected here so that the main text can focus on interpretation rather than table inventory.

\subsection{Dataset Cohorts}
\label{app:dataset_overviews}

\begin{table}[H]
\centering
\small
\caption{Overview of the boundary-scan dataset used in this work.}
\label{tab:dataset_boundary_overview}
\begin{tabular}{@{}p{0.22\linewidth}p{0.74\linewidth}@{}}
\toprule
Metric & Value \\
\midrule
Number of systems ($N$) & 765 \\
Basis sets & sto-3g (671), 6-31g (42), def2-svp (41), cc-pvdz (6), 6-31+g (3), aug-cc-pvdz (1), def2-svpd (1) \\
$n$ (qubits) & min=2, median=22, max=30 \\
$\mathrm{dim}(\mathcal{H})=2^n$ & $2^{2}$ to $2^{30}$ \\
Unique elements & 18 (Al, Ar, B, Be, C, Cl, F, H, He, Li, Mg, N, Na, Ne, O, P, S, Si) \\
R-parameter coverage & 455/765 (59.5\%); $R\in[0.50, 10.80]$, unique=184 \\
R sampling density & series=101, median pts/series=4.0, max=10; median $\Delta R$=0.47, min=0.12 \\
\bottomrule
\end{tabular}
\end{table}

\begin{table}[H]
\centering
\small
\caption{Overview of the performance-test dataset used in this work.}
\label{tab:dataset_performance_overview}
\begin{tabular}{@{}p{0.22\linewidth}p{0.74\linewidth}@{}}
\toprule
Metric & Value \\
\midrule
Number of systems ($N$) & 400 \\
Basis sets & sto-3g (389), 6-31g (11) \\
$n$ (qubits) & min=12, median=24, max=30 \\
$\mathrm{dim}(\mathcal{H})=2^n$ & $2^{12}$ to $2^{30}$ \\
Unique elements & 18 (Al, Ar, B, Be, C, Cl, F, H, He, Li, Mg, N, Na, Ne, O, P, S, Si) \\
R-parameter coverage & 272/400 (68.0\%); $R\in[0.75, 10.80]$, unique=161 \\
R sampling density & series=77, median pts/series=4.0, max=5; median $\Delta R$=0.53, min=0.13 \\
\bottomrule
\end{tabular}
\end{table}

\FloatBarrier

\clearpage
\subsection{Resource and Certificate Summaries}
\label{app:resource_certificate_overviews}

\begin{table}[H]
\centering
\scriptsize
\setlength{\tabcolsep}{1.5pt}
\renewcommand{\arraystretch}{0.95}
\caption{Resource-benefit summary at representative rank choices on the performance-test dataset.}
\label{tab:resource_benefit_overview}
\begin{tabular}{@{}c c r r r r@{}}
\toprule
Metric & Rank & $N$ & min & med & max \\
\midrule
$S$ & $k=1$ & 400 & $60.3$ & $1.65\times 10^{5}$ & $1.04\times 10^{6}$ \\
$S$ & $k^*_{0.999}$ & 347 & $32.7$ & $9.92\times 10^{3}$ & $5.32\times 10^{5}$ \\
$S$ & $k^*_{0.9995}$ & 236 & $4.33$ & $7.76\times 10^{3}$ & $1.03\times 10^{5}$ \\
$M$ & $k=1$ & 400 & $175$ & $3\times 10^{3}$ & $5.97\times 10^{3}$ \\
$M$ & $k^*_{0.999}$ & 347 & $9.61$ & $159$ & $2.3\times 10^{3}$ \\
$M$ & $k^*_{0.9995}$ & 236 & $2.24$ & $112$ & $974$ \\
$D$ & $k=1$ & 53 & $1.86$ & $19.7$ & $36.6$ \\
$D$ & $k^*_{0.999}$ & 39 & $0.31$ & $1.79$ & $14.8$ \\
$D$ & $k^*_{0.9995}$ & 19 & $0.0619$ & $0.929$ & $7.41$ \\
\bottomrule
\end{tabular}
\vspace{2pt}
\begin{minipage}{\linewidth}
\scriptsize\textit{Notes:} $S=t_{\mathrm{dense}}/t_{\mathrm{decomp}}$ (SVD-only speedup); $M=b_{\mathrm{dense}}/b_{\mathrm{factors}}$ (factor storage reduction); $D$ is an effective depth-saving factor on the circuit-evaluated subset, using the best cut (parallel depth) and scaled by $1/k$ to account for rank accumulation. $k^*_{\tau}$ denotes the smallest rank satisfying $\rho_k\ge\tau$ (reaching systems only).
\end{minipage}
\end{table}

\begin{table}[H]
\centering
\scriptsize
\setlength{\tabcolsep}{4pt}
\renewcommand{\arraystretch}{0.97}
\caption{Empirical validity and conservatism of the $\rho_k$-driven worst-case energy-error certificate.}
\label{tab:certificate_overview}
\begin{tabular}{@{}p{0.72\linewidth}p{0.20\linewidth}@{}}
\toprule
Metric & Value \\
\midrule
Total systems & 350 \\
Systems with observed errors & 24 \\
Evaluated (system, $k$) records & 768 \\
Bound pass rate & 768/768 (100.0\%) \\
Tightness ratio $\eta$ (median) & 0.01423 \\
Tightness ratio $\eta$ ($90$th pct.) & 0.02795 \\
Tightness ratio $\eta$ (max) & 0.03834 \\
Chemical accuracy $\epsilon_{\mathrm{chem}}$ & $1.59\times 10^{-3}$~Ha \\
\midrule
Fixed $\rho$ target reachability ($\rho\ge 0.999$) & 297/350 (84.9\%) \\
Fixed $\rho$ target reachability ($\rho\ge 0.9999$) & 32/350 (9.1\%) \\
Fixed $\rho$ target sufficient for chemical bound ($\rho\ge 0.999$) & 0/350 (0.0\%) \\
Fixed $\rho$ target sufficient for chemical bound ($\rho\ge 0.9999$) & 0/350 (0.0\%) \\
\bottomrule
\end{tabular}
\end{table}

\FloatBarrier

%
%
%
%
%
%
%

\end{document}